\documentclass[12pt]{article}

\usepackage[utf8]{inputenc}
\usepackage{textcomp}
\usepackage{chetnew}
\usepackage{tikz}
\usepackage{amssymb,amsfonts,mathrsfs,dsfont,yfonts,bbm}\usetikzlibrary{calc}
\usepackage{xcolor}
\usepackage{braket}
\usepackage{tikz-cd}
\usepackage{cite}

\newcommand{\trmax}{\rho_{\text{max}}}
\newcommand{\trep}{\rho}
\newcommand{\timess}{\times}
\newcommand{\co}[1]{{}^{#1}}

\newcommand{\be}{\begin{equation}}
\newcommand{\ee}{\end{equation}}
\newcommand{\bea}{\begin{eqnarray}}
\newcommand{\eea}{\end{eqnarray}}

\newcommand{\DZ}{\mathds{Z}}

\newcommand{\CC}{\mathcal{C}}

\newcommand{\CT}{\mathcal{T}}

\usepackage{amsmath}	
\begin{document}

\title{On Reconstructing Finite Gauge Group \\
from Fusion Rules}

\author{Rajath Radhakrishnan}
\affiliation{
International Centre for Theoretical Physics, Strada Costiera 11, Trieste 34151, Italy}

\abstract{Abstract: Gauging a finite group 0-form symmetry $G$ of a quantum field theory (QFT) results in a QFT with a Rep$(G)$ symmetry implemented by Wilson lines. The group $G$ determines the fusion of Wilson lines. However, in general, the fusion rules of Wilson lines do not determine $G$. In this paper, we study the properties of $G$ that can be determined from the fusion rules of Wilson lines and surface operators obtained from higher-gauging Wilson lines. This is in the spirit of Richard Brauer who asked what information in addition to the character table of a finite group needs to be known to determine the group. We show that fusion rules of surface operators obtained from higher-gauging Wilson lines can be used to distinguish infinite pairs of groups which cannot be distinguished using the fusion of Wilson lines. We derive necessary conditions for two non-isomorphic groups to have the same surface operator fusion and find a pair of such groups.}

\setcounter{tocdepth}{2}

\maketitle
\toc

\section{Introduction}

\begin{figure}[h!]
\centering

\tikzset{every picture/.style={line width=0.75pt}} 

\begin{tikzpicture}[x=0.75pt,y=0.75pt,yscale=-1,xscale=1]

\draw [color={rgb, 255:red, 139; green, 6; blue, 24 }  ,draw opacity=1 ][line width=0.75]    (211,81) -- (211,202) ;
\draw [color={rgb, 255:red, 139; green, 6; blue, 24 }  ,draw opacity=1 ][line width=0.75]    (280,81) -- (280,201) ;
\draw [color={rgb, 255:red, 139; green, 6; blue, 24 }  ,draw opacity=1 ][line width=0.75]    (433,81) -- (432,202) ;

\draw (173,65.4) node [anchor=north west][inner sep=0.75pt]    {$W_{R_{i}}$};
\draw (244,64.4) node [anchor=north west][inner sep=0.75pt]    {$W_{R_{j}}$};
\draw (311,139.4) node [anchor=north west][inner sep=0.75pt]    {$=$};
\draw (337,127.4) node [anchor=north west][inner sep=0.75pt]    {$\sum _{R_{k}} \ N_{R_{i} ,R_{j}}^{R_{k}}$};
\draw (398,66.4) node [anchor=north west][inner sep=0.75pt]    {$W_{R_{k}}$};

\end{tikzpicture}
\caption{Two topological Wilson lines can be brought together to form a single Wilson line with charge $R_i \times R_j$. The Wilson line $W_{R_i \times R_j}$ can be decomposed into simple Wilson lines $W_{R_k}$ where $R_i \times R_k= \sum_{R_k} N_{R_i R_j}^{R_k} R_k$.}
\label{fig:Wfusion}
\end{figure}
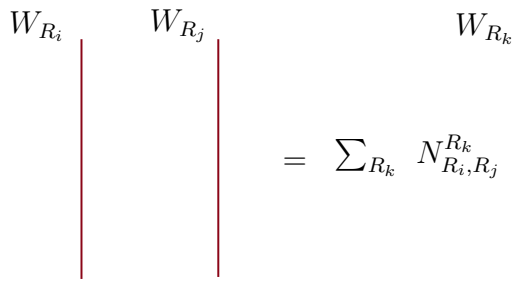

Various generalizations of the notion of symmetries have been an active area of research in recent years \cite{cordova2022snowmass}. For finite group symmetries, early works in this direction involve the study of orbifolding in string theory and 2-dimensional conformal field theories. One of the main lessons from this was that gauging a non-anomalous finite abelian 0-form symmetry $G$ of a QFT $\CT$ results in a QFT $\CT/G$ with a dual symmetry $\hat G=\text{Hom}(G,U(1))$ \cite{avafa1989quantum,gaiotto2015generalized}. Gauging the $\hat G$ symmetry of $\CT/G$ QFT results in the QFT $\CT$. $G$ and $\hat G$ are isomorphic as groups. This can be generalized to gauging a non-abelian group $G$ and the dual symmetry, in this case, is denoted Rep$(G)$ \cite{bhardwaj2018finite,bischoff2014tensor,brunner2015discrete,carqueville2016orbifold,etingof2016tensor}. Rep$(G)$ is the symmetry implemented by Wilson lines $W_{R}$ labelled by the representations $R$ of $G$. For non-abelian $G$, Rep$(G)$ is not a group, but rather a ring. The multiplication in the ring is given by the fusion of Wilson lines. For Wilson lines labelled by irreducible representations $R_i$ and $R_j$, fusion is given by the decomposition of the tensor product of representations into irreducible representations (see fig. \ref{fig:Wfusion}).
\be
W_{R_i} \times W_{R_j}= \sum_{R_k} N_{R_i R_j}^{R_k} W_{R_k}, ~~ N_{R_iR_j}^{R_k} \in \mathds{N}~.
\ee

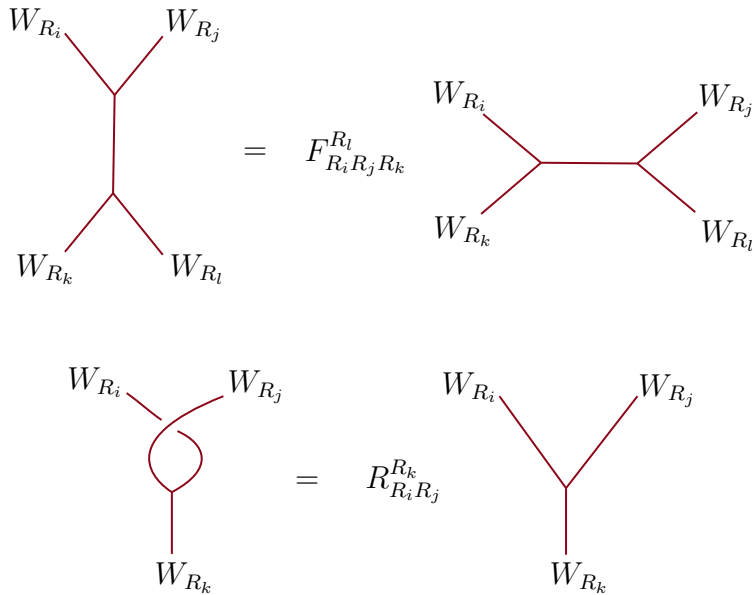
\begin{figure}[h!]
\centering

\tikzset{every picture/.style={line width=0.75pt}} 

\begin{tikzpicture}[x=0.75pt,y=0.75pt,yscale=-0.8,xscale=0.8]

\draw [color={rgb, 255:red, 139; green, 6; blue, 24 }  ,draw opacity=1 ]   (119.31,60.15) -- (150.66,99) ;
\draw [color={rgb, 255:red, 139; green, 6; blue, 24 }  ,draw opacity=1 ]   (181.02,62) -- (150.66,99) ;
\draw [color={rgb, 255:red, 139; green, 6; blue, 24 }  ,draw opacity=1 ]   (150.66,99) -- (149.68,161) ;
\draw [color={rgb, 255:red, 139; green, 6; blue, 24 }  ,draw opacity=1 ]   (149.68,161) -- (181.02,199.85) ;
\draw [color={rgb, 255:red, 139; green, 6; blue, 24 }  ,draw opacity=1 ]   (149.68,161) -- (119.32,198) ;
\draw [color={rgb, 255:red, 139; green, 6; blue, 24 }  ,draw opacity=1 ]   (381.71,174.1) -- (419.48,141.77) ;
\draw [color={rgb, 255:red, 139; green, 6; blue, 24 }  ,draw opacity=1 ]   (382.99,111.08) -- (419.48,141.77) ;
\draw [color={rgb, 255:red, 139; green, 6; blue, 24 }  ,draw opacity=1 ]   (419.48,141.77) -- (480.21,142.23) ;
\draw [color={rgb, 255:red, 139; green, 6; blue, 24 }  ,draw opacity=1 ]   (480.21,142.23) -- (517.98,109.9) ;
\draw [color={rgb, 255:red, 139; green, 6; blue, 24 }  ,draw opacity=1 ]   (480.21,142.23) -- (516.7,172.92) ;

\draw (81.66,42.4) node [anchor=north west][inner sep=0.75pt]    {$W_{R_{i}}$};
\draw (180.7,43.4) node [anchor=north west][inner sep=0.75pt]    {$W_{R_{j}}$};
\draw (86.55,197.4) node [anchor=north west][inner sep=0.75pt]    {$W_{R_{k}}$};
\draw (183.63,197.4) node [anchor=north west][inner sep=0.75pt]    {$W_{R_{l}}$};
\draw (518.34,176.32) node [anchor=north west][inner sep=0.75pt]    {$W_{R_{l}}$};
\draw (349.91,173.4) node [anchor=north west][inner sep=0.75pt]    {$W_{R_{k}}$};
\draw (349.04,88.4) node [anchor=north west][inner sep=0.75pt]    {$W_{R_{i}}$};
\draw (516.61,90.4) node [anchor=north west][inner sep=0.75pt]    {$W_{R_{j}}$};
\draw (229.74,130.4) node [anchor=north west][inner sep=0.75pt]    {$=$};
\draw (268.6,120.4) node [anchor=north west][inner sep=0.75pt]    {$F_{R_{i} R_{j} R_{k}}^{R_{l}}$};

\end{tikzpicture}

\vspace{1.0cm}

\tikzset{every picture/.style={line width=0.75pt}} 

\begin{tikzpicture}[x=0.75pt,y=0.75pt,yscale=-0.8,xscale=0.8]

\draw [color={rgb, 255:red, 139; green, 6; blue, 24 }  ,draw opacity=1 ]   (426.17,90.33) -- (468.17,148.33) ;
\draw [color={rgb, 255:red, 139; green, 6; blue, 24 }  ,draw opacity=1 ]   (468.17,148.33) -- (468.17,185.33) -- (468.17,190.33) ;
\draw [color={rgb, 255:red, 139; green, 6; blue, 24 }  ,draw opacity=1 ]   (219.67,150.83) -- (219.67,171.33) -- (219.67,189.83) ;
\draw [color={rgb, 255:red, 139; green, 6; blue, 24 }  ,draw opacity=1 ]   (219.67,150.83) .. controls (209.17,144.33) and (181.17,118.33) .. (252.17,90.33) ;
\draw [color={rgb, 255:red, 139; green, 6; blue, 24 }  ,draw opacity=1 ]   (191.17,88.33) .. controls (198.17,93.33) and (207.17,101.33) .. (213.17,105.33) ;
\draw [color={rgb, 255:red, 139; green, 6; blue, 24 }  ,draw opacity=1 ]   (223.17,112.33) .. controls (241.17,120.33) and (247.17,136.33) .. (219.67,150.83) ;
\draw [color={rgb, 255:red, 139; green, 6; blue, 24 }  ,draw opacity=1 ]   (513.17,90.33) -- (468.17,148.33) ;

\draw (344,144) node    {$=\ \ \ R_{R_{i} R_{j}}^{R_{k}}$};
\draw (152,70.4) node [anchor=north west][inner sep=0.75pt]    {$W_{R_{i}}$};
\draw (253,72.4) node [anchor=north west][inner sep=0.75pt]    {$W_{R_{j}}$};
\draw (207,193.4) node [anchor=north west][inner sep=0.75pt]    {$W_{R_{k}}$};
\draw (456,193.4) node [anchor=north west][inner sep=0.75pt]    {$W_{R_{k}}$};
\draw (388,73.4) node [anchor=north west][inner sep=0.75pt]    {$W_{R_{i}}$};
\draw (512,74.4) node [anchor=north west][inner sep=0.75pt]    {$W_{R_{j}}$};

\end{tikzpicture}
\caption{The crossing symmetry of Wilson lines along with their braiding define the matrices $F,R$ which along with $N_{R_i,R_j}^{R_k}$ can be used to reconstruct the gauge group $G$ up to isomorphism.}
\label{fig:FandR}
\end{figure}

The symmetry group $G$ of the QFT $\CT$ completely determines the fusion coefficients $N_{R_i,R_j}^{R_j}$. Since gauging the Rep$(G)$ symmetry of $\CT/G$ results in the QFT $\CT$ with $G$ symmetry, it is natural to wonder whether the ring structure/fusion coefficients of Rep$(G)$ determine the group $G$. This turns out to be not the case. For example, for the Dihedral group $D_8$ and Quaternion group $Q_8$ of order $8$, the fusion of Wilson lines are the same. In other words, the rings Rep$(D_8)$ and Rep$(Q_8)$ are isomorphic. There are many such pairs and even infinite families of non-isomorphic groups with the same Rep$(G)$. In fact, determining $G$ from Rep$(G)$ requires not just the fusion coefficients of Rep$(G)$, but also the full symmetric tensor category structure of Rep$(G)$ \cite{joyal1991introduction}. Physically, this means that along with the fusion rules, we also need the solutions to the crossing equations and braiding for Wilson lines to determine the group $G$. These are determined by the $F$ and $R$ matrices\footnote{The $R$ matrices are not to be confused with the labels $R_i$ for the irreducible representations of $G$. In the rest of the paper, we won't be using the $R$ matrices.} (see fig. \ref{fig:FandR}). Note that even the data $N_{R_iR_j}^{R_k}$ and $F$ do not fix the group $G$ up to isomorphism \cite{etingof2001isocategorical}. This is because a fusion category can admit different symmetric braidings leading to distinct $R$ matrices. Even though the group $G$ can be reconstructed from $N_{R_iR_j}^{R_k},F,R$, determining the $F$ and $R$ matrices is not trivial (see \cite{hu2013emergent} for a method to compute $F$ matrices for certain finite groups). Moreover, $F$ and $R$ matrices depend on a choice of basis for the Hilbert space of local operators at the trivalent junctions of Wilson lines. Therefore, it is not straightforward to compare two sets of $F,R$ data\footnote{The $S$ and $T$ matrices (which are independent of a basis) cannot be used to distinguish different groups as they are trivial for Rep$(G)$.}. This is unlike the fusion coefficients $N_{R_i,R_j}^{R_k}$ which do not depend on a basis. 

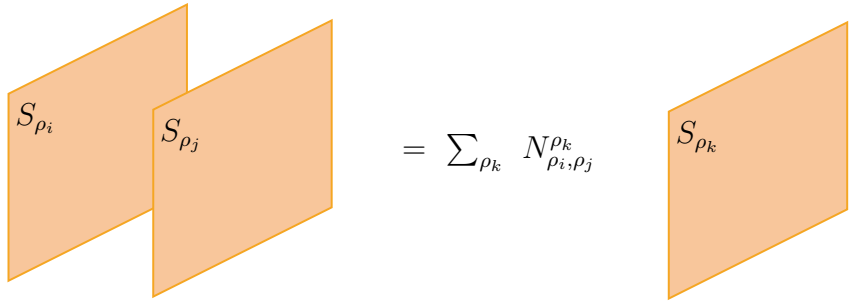
\begin{figure}[h!]
\centering

\tikzset{every picture/.style={line width=0.75pt}} 

\begin{tikzpicture}[x=0.75pt,y=0.75pt,yscale=-1,xscale=1]

\draw  [color={rgb, 255:red, 245; green, 166; blue, 35 }  ,draw opacity=1 ][fill={rgb, 255:red, 248; green, 198; blue, 155 }  ,fill opacity=1 ] (121.99,220.86) -- (122.01,126.49) -- (211.99,81.45) -- (211.98,175.82) -- cycle ;
\draw  [color={rgb, 255:red, 245; green, 166; blue, 35 }  ,draw opacity=1 ][fill={rgb, 255:red, 248; green, 198; blue, 155 }  ,fill opacity=1 ] (194.99,228.86) -- (195.01,134.49) -- (284.99,89.45) -- (284.98,183.82) -- cycle ;
\draw  [color={rgb, 255:red, 245; green, 166; blue, 35 }  ,draw opacity=1 ][fill={rgb, 255:red, 248; green, 198; blue, 155 }  ,fill opacity=1 ] (454.99,229.86) -- (455.01,135.49) -- (544.99,90.45) -- (544.98,184.82) -- cycle ;

\draw (124.01,129.89) node [anchor=north west][inner sep=0.75pt]    {$S_{\rho _{i}}$};
\draw (319,151.4) node [anchor=north west][inner sep=0.75pt]    {$=$};
\draw (341,145.4) node [anchor=north west][inner sep=0.75pt]    {$\sum _{\rho _{k}} \ N_{\rho _{i} ,\rho _{j}}^{\rho _{k}}$};
\draw (197.01,137.89) node [anchor=north west][inner sep=0.75pt]    {$S_{\rho _{j}}$};
\draw (457.01,138.89) node [anchor=north west][inner sep=0.75pt]    {$S_{\rho _{k}}$};
\end{tikzpicture}

\caption{Two topological surface operators can be brought together to form a single surface operator $S_{\trep_i \times \trep_j}$. $S_{\trep_i \times \trep_j}$ can be decomposed into simple surface operators $S_{\trep_k}$ where $\trep_i \times \trep_k= \sum_{\trep_k} N_{\trep_i \trep_j}^{\trep_k} \trep_k$.}
\label{fig:Sfusion}
\end{figure}

\textit{Can we add other basis-independent data to $N_{R_i,R_j}^{R_k}$ to determine the group $G$?} In particular, what do the fusion rules of other operators in the QFT tell us about $G$? To answer this question, let us assume that the QFT $\CT$ is at least 3-dimensional. Consider surface operators $S_{\trep}$ of $\CT/G$ obtained by higher gauging the Rep$(G)$ symmetry on 2D submanifolds of the spacetime \cite{roumpedakis2022higher,bhardwaj2022universal,bartsch2022non}\footnote{Through out this paper, the term `surface operator' will refer to the surface operators obtained from higher-gauging the Wilson lines. The QFT $\CT/G$ has many other surface operators. But their fusion depends on the QFT $\CT$ we start with and will not play a role in our discussion.}. Like the Wilson lines, the surface operators $S_{\trep}$ are labelled by certain representations $\trep$ of the group $G$. More precisely, $\rho$ is a 2-representation of the group $G$ \cite{bartsch2022non}. For two surface operators $S_{\trep_i}$ and $S_{\trep_j}$ labelled by irreducible 2-representations $\trep_i,\trep_j$, their fusion is given by the decomposition of the tensor product of 2-representations into irreducible 2-representations (see fig. \ref{fig:Sfusion}).
\be
S_{\trep_i} \times S_{\trep_j}= \sum_k N_{\trep_i, \trep_j}^{\trep_k} S_{\trep_k}, ~~ N_{\trep_i, \trep_j}^{\trep_k}\in \mathds{N}~.
\ee

The fusion of surface operators endows the set of surface operators with a ring structure, which we will denote as 2Rep$(G)$. In summary, the Wilson lines $W_R$ and surface operators $S_\trep$ implement non-invertible symmetries of $\CT/G$ given by the rings Rep$(G)$ and 2Rep$(G)$, respectively\footnote{In general, only a subset of these lines and surfaces act faithfully on the QFT.}.  

The symmetry group $G$ of the QFT $\CT$ completely determines the fusion coefficients $N_{R_i R_j}^{R_k}$ and $N_{\trep_i, \trep_j}^{\trep_k}$. {\it What can we learn about the group $G$ given the data $N_{R_i R_j}^{R_k}$ and $N_{\trep_i, \trep_j}^{\trep_k}$?} As we will show in the bulk of the paper, the groups $D_8$ and $Q_8$ which cannot be distinguished by the fusion of Wilson lines \textit{can} be distinguished using the fusion of surface operators. More generally, we will study the properties of the group $G$ that can be determined from the fusion rules of Wilson lines and surface operators. 
We will use these properties to show that infinite pairs of groups which cannot be distinguished using the fusion of Wilson lines can be distinguished using the fusion of surface operators. We will derive necessary conditions for two non-isomorphic groups to have the same surface operator fusion. Using this, we will show that all groups of order $< 96$ can be distinguished using $N_{\trep_i, \trep_j}^{\trep_k}$. On the other hand, by studying an explicit example, we will show that non-isomorphic groups can have the same surface operator fusion. 

Before explaining the structure of the paper, let us briefly consider QFTs in 4 dimensions or higher. Gauging a $G$ symmetry of such a QFT $\CT$, we get a QFT $\CT/G$ with Wilson lines, condensation surfaces as well as condensation membrane operators \cite{Bartsch:2022ytj,bhardwaj2022universal}. These are labelled by representations, 2-representations and 3-representations of the group $G$, respectively \cite{Bartsch:2022ytj}. From the results in this paper, it is natural to expect that the fusion coefficients of Wilson lines, surface operators and membrane operators together can be used to derive even more properties of the group $G$. This requires a study of the tensor product of 3-representations which we leave for future work.

The structure of the paper is as follows. In section \ref{sec:Wilson} we describe the fusion of Wilson lines and relate to it the character table of the gauge group $G$. We study various properties of $G$ that can be determined from Wilson lines. We end this section by studying the groups $D_8$ and $Q_8$ which have isomorphic Wilson line fusion. In section \ref{sec:S2rep} we review 2-representation theory of a finite group. We end this section by explicitly computing the fusion rules of surface operators corresponding to $G=\DZ_6, S_3, D_8$ and $Q_8$. We continue studying the properties of fusion of surface operators in section \ref{sec:GfromSfus}. We describe how to deduce various properties of $G$ from these fusion rules. A summary of different types of fusion rules and the properties of $G$ that can be deduced from them is given in fig. \ref{fig:ressummary}. We study infinite pairs of groups with the same Wilson line fusion and show that these groups can be distinguished using surface operator fusion. We end this section by deriving the necessary conditions for two groups to have isomorphic surface operator fusion. We use these conditions to identify a pair of such groups. Finally, in section \ref{sec:GfromLonS} we study the fusion rules of non-genuine line operators and show that it determines $G$ up to isomorphism. We conclude with a summary of the results and a list of directions for future research.  

Appendix \ref{Ax:2char} reviews the character theory for 2-representations and describes the results in the bulk of the paper in terms of 2-characters.

\section{$G$ from fusion of Wilson lines}
\label{sec:Wilson}

Gauging a 0-form finite group symmetry $G$ of a QFT $\CT$ results in a QFT $\CT/G$ with  symmetry Rep$(G)$ implemented by Wilson lines. As we mentioned in the introduction, the fusion of Wilson lines is given by
\be
W_{R_i} \times W_{R_j}= \sum_i N_{R_iR_j}^{R_k} W_{R_k}~.
\ee
The fusion coefficients $N_{R_i,R_j}^{R_k}$ contain the same information about the group as in the character table. Indeed, the full character table of the group can be constructed by computing eigenvalues of the matrices $(\hat N_{R_i})_{R_j,R_k}:=N_{R_i,R_j}^{R_k}$ (see \cite{adrian2023rowcol} for an explicit construction). Conversely, we can determine $N_{R_i,R_j}^{R_k}$ from the characters using the formula
\be
N_{R_i,R_j}^{R_k}= \frac{1}{|G|} \sum_{C \in Cl(G)} |C|~ \chi_{R_i}(C)\chi_{R_j}(C)\chi_{R_k}(C)^{-1}~,
\ee
where $C$ is a conjugacy class of $G$ and $Cl(G)$ is the set of all conjugacy classes of $G$. In this formula $\chi_{R_i}(C)$ is the evaluation of the character on a representative of the conjugacy class $C$. Therefore, all properties of $G$ that can be determined from the characters can also be determined from the fusion of Wilson lines. Let us label the rows of the character table by the irreducible representations and the columns by conjugacy classes. The sum of squares of dimensions of the irreducible representations gives us the order of the group
\be
|G|=\sum_R (\chi_R(1))^2~.
\ee
More generally, the sum along a column of the character table gives
\be
|N_{C}|:= \sum_R |\chi_R(C)|^2~,
\ee
where $|N_C|$ is the size of the centralizer of a representative of $C$. The normal subgroups of the group $G$, their orders and intersections can be determined using the character table. $N$ is a normal subgroup of $G$ if and only if there exists a character $\chi_{R}$ for some (possible reducible) representation $R$ such that \cite{isaacs2006character}
\be
N=\text{ker}(\chi_R):=\{C \in Cl(G)~|~\chi_R(C)=|R|\}~,
\ee
where $|R|$ is the dimension of the representation $R$. In particular, we get some normal subgroups $N_i$ 
\be
N_i=\text{ker}(\chi_{R_i})~,
\ee
from the irreducible representations $R_i$. All normal subgroups can be obtained by taking intersections of the $N_i$. Of course, if we don't have the group $G$ to begin with we don't know the group multiplication and subgroups of $G$. Therefore, by determining a normal subgroup $N\vartriangleleft G$ from the character table, what we mean is that we can identify $N$ as a union of conjugacy classes of $G$. In particular, the modular lattice of normal subgroups can be determined from the character table. The character table of $G$ also determines the character table of the group $G/N$ for any normal subgroup $N$ in $G$. For more details, see \cite{brauer1963representations,isaacs2006character}. 
If the group is simple, then using the classification of simple groups, it can be shown that the character table determines the simple group uniquely.

Even though the fusion rules of Wilson lines (equivalently the character table) contain a lot of information about the group, as is well known, it does not uniquely determine a group. The character table of $G$ does not determine the character table of $N \triangleleft G$. Moreover, even though it determines the size of all normal subgroups, the isomorphism class of the normal subgroups cannot be always determined. It cannot be determined whether a given normal subgroup is abelian or not. For more details, see \cite{ mattarei1992retrieving,mattarei1994example}. In general, non-isomorphic groups can have isomorphic character tables (see \cite{cocke2019database} for a database of such groups). Groups of the smallest order which share the same character table are the Dihedral group $D_8$ and the Quaternion group $Q_8$ of order $8$. To understand and contrast with the discussion in the following sections, it is useful to study this example in more detail.  $D_8$ group has the presentation
\be
D_8:=\langle r,s | r^{4}=s^2=(sr)^2=1\rangle~.
\ee
The conjugacy classes of this group are
\be
[1], [r^2], [r], [s], [rs]~.
\ee
where $[c]$ denotes the conjugacy class with representative $c$. The character table is
\be
\begin{tabular}{c|ccccc}
    & [1] & [$r^2$] & [r] & [$s$] & [$rs$] \\
\hline
$R_1$ &  1 & 1 & 1 & 1 & 1\\
$R_2$ & 1 & 1 & 1 & -1 &  -1 \\
$R_3$ & 1 & 1 & -1 & 1 & -1 \\
$R_4$ & 1 & 1 & -1 & -1 & 1 \\
$R_5$ & 2 & -2 & 0 & 0 & 0
\end{tabular}
\ee
We can identify the normal subgroups of $D_8$ by looking at the kernels of the characters in the table above. We find the following normal subgroups
\be
D_4,~ \{[1],[r],[r^2]\}\simeq \DZ_4,~ \{[1],[r^2],[s]\}\simeq \DZ_2 \times \DZ_2,~ \{[1],[r^2],[rs]\} \simeq \DZ_2 \times \DZ_2,~ \{[1]\} ~.
\ee
All other normal subgroups of $D_8$ are intersections of these normal subgroups. Indeed, we get $\{1,r^2\}\simeq \DZ_2$ which exhausts all the normal subgroups of $D_8$. The isomorphism class of the normal subgroups is not determined by the character table. In general, the non-normal subgroups of a group cannot be determined from the character table. The non-normal subgroups of $D_8$ are $\{1,s\},\{1,rs\},\{1,r^2s\},\{1,r^3s\}$ which are all isomorphic to $\DZ_2$.

$Q_8$ group has the presentation
\be
Q_8:=\langle i,j,k | i^2=j^2=k^2=ijk, (ijk)^2=1 \rangle~.
\ee
The conjugacy classes of this group are
\be
[1], [ijk], [i], [j], [k]~.
\ee
The character table is
\be
\begin{tabular}{c|ccccc}
    & [1] & [$ijk$] & [i] & [$j$] & [$k$] \\
\hline
$R_1$ &  1 & 1 & 1 & 1 & 1\\
$R_2$ & 1 & 1 & 1 & -1 &  -1 \\
$R_3$ & 1 & 1 & -1 & 1 & -1 \\
$R_4$ & 1 & 1 & -1 & -1 & 1 \\
$R_5$ & 2 & -2 & 0 & 0 & 0
\end{tabular}
\ee
We can identify the normal subgroups of $Q_8$ by looking at the kernels of the characters in the table above. We find the following normal subgroups
\be
Q_4, \{[1],[ijk],[i]\}\simeq \DZ_4, \{[1],[ijk],[j]\}\simeq \DZ_4, \{[1],[ijk],[k]\} \simeq \DZ_4, \{[1]\} ~.
\ee
All other normal subgroups of $Q_8$ are subgroups of these normal subgroups. Indeed, we get $\{1,ijk\}\simeq \DZ_2$ which exhausts all the normal subgroups of $Q_8$. Since all subgroups of $Q_8$ are normal, all subgroups of $Q_8$ can be determined from the character table.  

Since $D_8$ and $Q_8$ have the same character table, the fusion of Wilson lines clearly cannot distinguish between $D_8$ and $Q_8$. What extra data on top of the fusion of Wilson lines do we need to distinguish between $Q_8$ and $D_8$? This is equivalent to asking about the information that needs to be known in addition to the character table to determine the group. In finite group theory, this is a crucial question asked by Richard Brauer in \cite{brauer1963representations}. One way forward is to use the crossing symmetry of Wilson lines and compute the $F$ matrices. The $F$ matrices for $D_8$ and $Q_8$ are distinct (see \cite{hu2013emergent}, \cite{etingof2001isocategorical} and \cite{11346}). However, the $F$ matrices are basis dependent and hard to compute in general. It is natural to wonder if we can distinguish between $D_8$ and $Q_8$ using basis-independent data. In particular, can we distinguish between these groups using fusion rules of other operators in the QFT $\CT/G$.  In the following sections, we will show that surface operator fusion contains information about non-normal subgroups of the gauge group which can be used to distinguish between $D_8$ and $Q_8$, and other infinite pairs of groups with the same character table.

\section{Surface operators in $G$-gauge theory}

\label{sec:S2rep}

\begin{figure}[h!]
    \centering

\tikzset{every picture/.style={line width=0.75pt}} 

\begin{tikzpicture}[x=0.75pt,y=0.75pt,yscale=-1,xscale=1]

\draw  [draw opacity=0][fill={rgb, 255:red, 255; green, 236; blue, 204 }  ,fill opacity=1 ] (12.06,103.65) -- (54.96,60.75) -- (226.06,60.75) -- (226.06,160.85) -- (183.16,203.75) -- (12.06,203.75) -- cycle ; \draw  [draw opacity=0] (226.06,60.75) -- (183.16,103.65) -- (12.06,103.65) ; \draw  [draw opacity=0] (183.16,103.65) -- (183.16,203.75) ;
\draw  [color={rgb, 255:red, 245; green, 166; blue, 35 }  ,draw opacity=1 ][fill={rgb, 255:red, 248; green, 198; blue, 155 }  ,fill opacity=1 ] (140.09,61.51) -- (141.99,158.74) -- (96.57,203.29) -- (94.67,106.06) -- cycle ;
\draw  [draw opacity=0][fill={rgb, 255:red, 255; green, 236; blue, 204 }  ,fill opacity=1 ] (431.06,103.65) -- (473.96,60.75) -- (645.06,60.75) -- (645.06,160.85) -- (602.16,203.75) -- (431.06,203.75) -- cycle ; \draw  [draw opacity=0] (645.06,60.75) -- (602.16,103.65) -- (431.06,103.65) ; \draw  [draw opacity=0] (602.16,103.65) -- (602.16,203.75) ;
\draw  [color={rgb, 255:red, 245; green, 166; blue, 35 }  ,draw opacity=1 ][fill={rgb, 255:red, 248; green, 198; blue, 155 }  ,fill opacity=1 ] (559.1,61.74) -- (560.99,158.74) -- (515.59,203.36) -- (513.7,106.35) -- cycle ;
\draw    (285,139) -- (372,139) ;
\draw [shift={(374,139)}, rotate = 180] [color={rgb, 255:red, 0; green, 0; blue, 0 }  ][line width=0.75]    (10.93,-3.29) .. controls (6.95,-1.4) and (3.31,-0.3) .. (0,0) .. controls (3.31,0.3) and (6.95,1.4) .. (10.93,3.29)   ;

\draw (56.96,64.15) node [anchor=north west][inner sep=0.75pt]    {$\mathcal{T}$};
\draw (475.96,64.15) node [anchor=north west][inner sep=0.75pt]    {$\mathcal{T} /G$};
\draw (125,73.4) node [anchor=north west][inner sep=0.75pt]    {$\mathcal{C}$};
\draw (545,74.4) node [anchor=north west][inner sep=0.75pt]    {$S$};
\draw (251,115.4) node [anchor=north west][inner sep=0.75pt]    {\text{Diagonal $G$-gauging}};
\end{tikzpicture} 
\caption{Coupling the $G$-symmetric TQFT $\CC$ to the QFT $\CT$ and gauging the diagonal $G$-symmetry results in a surface defect $S$ in $\CT/G$.}
    \label{fig:diagGgauge}
\end{figure}
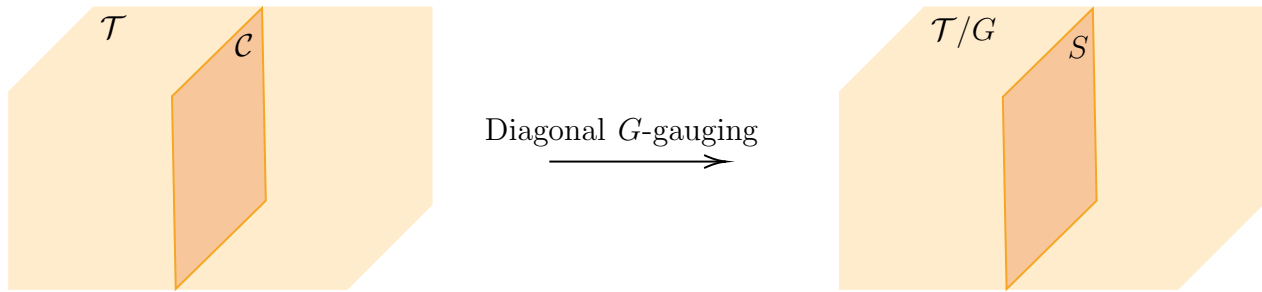

Consider a $d\geq 3$-dimensional quantum field theory $\CT$ with a non-anomalous finite group symmetry $G$. The QFT $\CT/G$ has Wilson lines which form the ring Rep$(G)$. Rep$(G)$ is a non-invertible symmetry of $\CT/G$ which can be gauged to get back $\CT$. Instead of gauging Rep$(G)$ on the full $d$-dimensional spacetime, we can higher-gauge Rep$(G)$ on a 2-dimensional surface to obtain a surface operator \cite{roumpedakis2022higher,bartsch2022non,bhardwaj2022universal}. Moreover, subcategories of Rep$(G)$ can be higher-gauged to define other surface operators. Another perspective on these surface operators is obtained by considering $G$-symmetric 2D TQFTs. Consider the QFT $\CT$ with 0-form symmetry $G$. We can couple this theory to a G-symmetric 2D TQFT $\CC$.
Gauging the diagonal $G$ symmetry of this system results in the QFT $\CT/G$, and $\CC$ becomes a surface operator in $\CT/G$ (see fig. \ref{fig:diagGgauge}). We can classify such surface operators by classifying $G$-symmetric 2D TQFTs  \cite{bhardwaj2022universal}. 

Wilson lines are labelled by representations of $G$. It is natural to wonder whether the surface operators described above are labelled by representation theoretic quantities of the group $G$. In \cite{bartsch2022non}, the authors show that these surface operators are described by 2-representations of $G$. The fusion of surface operators is given by the tensor product of 2-representations. In the following subsections, we will follow the description of surface operators given in \cite{bartsch2022non} in terms of 2-representations of $G$ and their tensor product. Determining $G$ from the fusion of surface operators/2-representations is a natural generalisation of the problem of determining $G$ from Wilson lines/1-representations.

\subsection{2-Representations of a Group}

Before describing 2-representations, let us define 2-matrices. An $n \times n$ 2-matrix $M$ is a matrix with vector space entries. That is, every $M_{ij}$ $1\leq i,j\leq n$ is a vector space. Matrix multiplication is defined using direct sum and tensor product of vector spaces as
\be
(M \times W)_{ij} := \bigoplus_k M_{ik} \otimes W_{kj}~,
\ee
where $\oplus$ and $\otimes$ denote the direct sum and tensor product of vector spaces, respectively. Note that 2-matrix multiplication is associative only up to isomorphisms
\be
M \times (W \times K) \simeq (M \times W) \times K~.
\ee

Recall that a complex representation (1-representation) of a group $G$ is a homomorphism $R$ from the group to the group of invertible matrices over the complex numbers. This homomorphism assigns a matrix $R(g)$ with every group element $g \in G$. Fix an integer $n$ (dimension of the 2-representation), a 2-representation $\rho$ assigns an $n \times n$ 2-matrix $\rho(g)$ with every group element $g\in G$ \cite{elgueta2004representation,ganter2008representation,osorno2010explicit}. In a 1-representation of a group, we require $R(g)R(h)=R(gh)$. Similarly, for 2-representations we require isomorphisms
\be
\phi_{g,h}: \rho(g)\times \rho(g) \rightarrow \rho(gh), ~ \phi_{1}: \rho(1) \rightarrow \mathds{1}_n ~,
\ee
where $1$ is the identity of the group and $\mathds{1}_n$ is the 2-matrix with 1-dimensional vector spaces along the diagonal and all other components are the zero vector space. We can use the isomorphisms $\phi_{g,h}$ to go from the 2-matrix product $\rho(g)\times \rho(h)\times \rho(k)$ to $\rho(ghk)$ in two different ways 
\be \label{eq:commut}
    \begin{tikzcd}[row sep=40pt, column sep=60pt]
        \displaystyle \rho(g)\times \rho(h)\times \rho(k) \arrow[r,"\displaystyle \text{Id}_{\rho(g)} \circ \phi_{h,k}",outer sep = 2pt] \arrow[d,"\displaystyle \phi_{g,h} \circ \text{Id}_{\rho(k)}"',outer sep = 3pt] & \displaystyle \rho(g) \times \rho(hk) \arrow[d, "\displaystyle \phi_{g,hk}", outer sep = 3pt] \\
        \displaystyle \rho(gh) \times \rho(k) \arrow[r,"\displaystyle \phi_{g,hk}"', outer sep = 3pt] & \displaystyle \rho(ghk)
    \end{tikzcd}
\ee
We require this diagram to commute. 

The solutions to these constraints are straightforward. The isomorphism $\rho(g)\rho(g^{-1})\simeq \mathds{1}_n$ implies that $\rho(g)$ is an $n \times n$ permutation 2-matrix. That is, $\rho(g)$ is a matrix with only one non-zero entry in every row and column. Moreover, this non-zero entry is a 1-dimensional vector space. By replacing every non-zero entry in $\rho(g)$ with the integer $1$ we get a permutation matrix. Let $\sigma_g$ denote this permutation. The isomorphisms $\phi_{g,h}$ are completely specified by $n$ phases $\tilde c_i(g,h)$, $1\leq i\leq n$. These phases should satisfy the constraint \eqref{eq:commut} which can be written as
\be
\label{eq:coh}
\tilde c_{\sigma^{-1}_g(i)}(h,k) \tilde c_i(g,hk)= \tilde c_i(g,h)\tilde c_i(gh,k)~.
\ee
Let $\tilde c(g,h)$ denote the set of $n$ phases $\tilde c_i(g,h)$. Then the constraint \eqref{eq:coh} implies that $\tilde c(g,h)$ is a 2-cocycle in the group $Z^2_{\sigma}(G,U(1)^n)$ where $G$ acts on $U(1)^n$ by permuting the $U(1)$ factors. Therefore, every 2-representation of the group $G$ is labelled by a permutation $\sigma_n$ and a 2-cocycle $\tilde c(g,h)$. In fact, by studying induced representations, the irreducible 2-representations can be shown to be labelled by the choice of a subgroup $K \subset G$ and a cohomology class $[c]\in H^2(K,U(1))$, up to conjugation \cite{ganter2008representation}. Therefore, in the rest of this article, we will label irreducible 2-representations by the tuple $(K,[c])$, where the notation $[c]$ indicates the cohomology class of the 2-cocycle $c$. Two irreducible two representations $(K,[c])$ and $(H,[d])$ are isomorphic if 
\be
H=\co{g}K, ~ [d]=[\co{g}c]~,
\ee
where $\co{g}K$ denotes the conjugation of elements of $K$ by $g\in G$ and $\co{g}c:=c(gag^{-1},gbg^{-1})$. Therefore, isomorphism classes of irreducible 2-representations are labelled by conjugacy classes of subgroups of $G$ and the cohomology classes of corresponding 2-cocycles. We will denote the set of 2-representations of a group $G$ by 2Rep$(G)$\footnote{2Rep$(G)$ is in fact a 2-category.}. 

\subsection{Fusion rules of surface operators}

In \cite{bartsch2022non}, the authors show that simple surface operators obtained from higher-gauging Wilson lines on a surface are labelled by irreducible 2-representations. We will label these surface operators as $S_{(K,[c])}$ where $(K,[c])$ is an irreducible 2-representation of $G$. The fusion of two surface operators $S_{(K,[c])}$ and $S_{(H,[d])}$ labelled by the irreducible 2-representations $(K,[c])$ and $(H,[d])$ of $G$ is given by the decomposition of tensor product of 2-representations into irreducible 2-representations \cite{bartsch2022non,greenough2010monoidal}
\be
\label{eq:surfacefusion}
S_{(K,[c])} \times S_{(H,[d])}= \sum_{g\in K\backslash G/H} S_{(K\cap\,\co{g}H,[c \cdot \,\co{g}d])}~.
\ee
where $[c \cdot \,\co{g}d]$ implicitly denotes the cohomology class of the 2-cocycle $c \cdot \,\co{g}d$ restricted to the group $K \cap \,\co{g}H$. We can rewrite this expression as
\be
\label{eq:surfacefusionN}
S_{(K,[c])} \times S_{(H,[d])}= \sum_{(L,[e])} N_{(K,[c]),(H,[d])}^{(L,[e])} ~ S_{(L,[e])}~,
\ee
where the integers $N_{(K,[c]),(H,[d])}^{(L,[e])}$ are defined as
\be
\label{eq:fusioncoef}
N_{(K,[c]),(H,[d])}^{(L,[e])}:= |\{ g\in K\backslash G/H ~|~(L,[e])=(K\cap \,\co{g}H,[c \cdot \,\co{g}d])\}|
\ee
The fusion rules of surface operators labelled by two representations satisfy
\bea
S_{(K,[c])} \times S_{(H,[d])}&=& \sum_{g\in K\backslash G/H} S_{(K\cap \,\co{g}H,[c \cdot \,\co{g}d])} = \sum_{g\in K\backslash G/H} S_{(\co{g^{-1}}K \cap H,[\co{g^{-1}}c \cdot d])} \\
&=& \sum_{g^{-1 }\in H\backslash G/K} S_{(\co{g^{-1}}K \cap H,[\co{g^{-1}}c \cdot d])}= S_{(H,[d])} \otimes S_{(K,[c])}~,
\eea
where in the first equality above we used the fact that the 2-representation labelled by $(K\cap \,\co{g}H,[c \cdot \,\co{g}d])$ and $(\co{g^{-1}}K\cap H,[\co{g^{-1}}c \cdot d])$ are equivalent. In the second equality above we used the isomorphism between the double cosets $g \in K\backslash G/H \to g^{-1} \in H\backslash G/K$. In other words, the fusion of condensation defects arising from higher gauging Rep$(G)$ is commutative. Note that the fusion of general condensation defects can be non-commutative \cite{roumpedakis2022higher}. 

From the fusion rules, we find that the trivial surface operator is $S_{(G,[1])}$. $S_{(K,[c])}$ is an invertible surface operator if and only if there exists another surface operator $S_{(H,[d])}$ such that
\be
S_{(K,[c])} \times S_{(H,[d])}= S_{(G,[1])}~.
\ee
Using \eqref{eq:surfacefusion} we find that invertible surface operators are precisely those labelled by the 2-representations $(G,[c])$ where $[c]\in H^2(G,U(1))$. Therefore, the group of invertible surface operators is $H^2(G,U(1))$. This shows that any $d\geq 3$-dimensional QFT with a Rep$(G)$ symmetry implemented by line operators contains invertible condensation surfaces implementing a $H^2(G,U(1))$ $(d-3)$-form symmetry. Note that these surfaces need not have a faithful action on the QFT.

The fusion of surface operators of the form $S_{(K,[1])}$ where $[1]$ denotes the trivial element of $H^2(K,U(1))$ is given by
\be
S_{(K,[1])} \times S_{(H,[1])}= \sum_{g\in K\backslash G/H} S_{(K\cap \,\co{g}H,[1])}~.
\ee
This is the Burnside ring $B(G)$ of the group $G$ \cite{bouc2000burnside}. $B(G)$ is the analogue of the representation ring (Rep$(G)$) of the group $G$ but now for representations of $G$ on finite sets. The fusion of Wilson lines and surface operators in a G-gauge theory gives a unified perspective on representations of $G$ on vector spaces and finite sets. For our discussion in the subsequent sections, it is useful to introduce the table of marks of a finite group. The table of marks of a group $G$ has rows and columns indexed by the conjugacy classes of subgroups of $G$. The entries of the table are given by
\be
m_{K,H}:=|\{ g\in K\backslash G/H ~|~ K\subseteq \,\co{g}H\}|~.
\ee
The fusion coefficients $N_{(K,[1]),(H,[1])}^{(L,[1])}$ can be written in terms of the table of marks by the formula
\be
N_{(K,[1]),(H,[1])}^{(L,[1])}= \sum_M m_{K,M}\, m_{H,M}\, m_{M,L}^{-1}~.
\ee
where the sum is over the conjugacy class of subgroups of $G$. Let $CS(G)$ be the set of conjugacy classes of subgroups of $G$. Two groups $G_1$ and $G_2$ are said to have isomorphic Burnside rings if there exists a bijection of sets $\sigma: CS(G_1) \to CS(G_2)$ such that
\be
N_{(K,[1]),(H,[1])}^{(L,[1])}=N_{(\sigma(K),[1]),(\sigma(H),[1])}^{(\sigma(L),[1])}~.
\ee
Similarly, two groups $G_1$ and $G_2$ are said to have isomorphic table of marks if there exists a bijection of sets $\sigma: CS(G_1) \to CS(G_2)$ such that
\be
m^{(G_1)}_{K,H}=m^{(G_2)}_{\sigma(K),\sigma(H)}~.
\ee
Two groups with isomorphic table of marks have isomorphic Burnside rings. An isomorphism of Burnside rings preserving the standard basis of conjugacy classes of subgroups implies an isomorphism of table of marks. 

The fusion of surface operators labelled by 2-representations of $G$ is a generalization of the Burnside ring to a twisted Burnside ring defined using $H^2(K,U(1))$, $K \subseteq G$ \cite{greenough2010monoidal}\footnote{Generalized Burnside rings based on $H^n(K,U(1))$ and associated table of marks are studied in \cite{hartmann2007generalized}.}. In the context of condensation defects, the relation between fusion rules of surface operators and twisted Burnside rings was explained in \cite{bhardwaj2022universal}. Similar to the definitions above, in section \ref{sec:isosfusion}, we will find groups with isomorphic surface operator fusion by defining an isomorphism of twisted Burnside rings and studying the necessary conditions for such an isomorphism to exist. 

\subsection{Examples}

\label{sec:examples}

\subsubsection{$\DZ_6$}

The group $\DZ_6$ has subgroups isomorphic to $\DZ_1,\DZ_2,\DZ_3,\DZ_6$. All of these subgroups have trivial $2^{\text{nd}}$ cohomology group. Therefore, we have 4 irreducible 2-representations labelled by $(\DZ_1,[1]),(\DZ_2,[1]),(\DZ_3,[1]),(\DZ_6,[1])$. The fusion rules for surface operators labelled by these representations are given by 
\begin{alignat}{4}
&S_{(\DZ_1,[1])} \times S_{(\DZ_1,[1])}= 6 S_{(\DZ_1,[1])}~,& ~
S_{(\DZ_1,[1])} \times S_{(\DZ_2,[1])}= 3 S_{(\DZ_1,[1])}~,\cr
&S_{(\DZ_1,[1])} \times S_{(\DZ_3,[1])}= 2 S_{(\DZ_1,[1])}~,& ~
S_{(\DZ_2,[1])} \times S_{(\DZ_2,[1])}= 3 S_{(\DZ_2,[1])}~,\\
&S_{(\DZ_2,[1])} \times S_{(\DZ_3,[1])}= S_{(\DZ_1,[1])}~,& ~
S_{(\DZ_3,[1])} \times S_{(\DZ_3,[1])}= 2 S_{(\DZ_3,[1])}~.\nonumber
\end{alignat}{4}
$S_{(\DZ_6,[1])}$ is the trivial surface operator. Note that the non-invertible surface operators $S_{(\DZ_2,[1])}$ and $S_{(\DZ_3,[1])}$ fuse to give a unique outcome\footnote{Non-invertible line operators can also have such fusions even if the gauge group is not a direct product \cite{Buican:2020who,buican2021non}.}. This fusion arises from the direct product structure of the group $\DZ_6 \simeq \DZ_2 \times \DZ_3$. In Section \ref{sec:groupext}, we will show that the fusion rules of surface operators can be used to determine whether the group is a direct product of smaller groups. 

\subsubsection{$S_3$}

Up to conjugation, $S_3$ group has the subgroups $\DZ_1,\DZ_2=\{(),(23)\},\DZ_3=\{(123),(132)\},S_3$. All of these groups have trivial $2^{\text{nd}}$ cohomology group. Therefore, the irreducible 2-representations are labelled by $(\DZ_1,[1]),(\DZ_2,[1]),(\DZ_3,[1]),(S_3,[1])$. We have the same number of 2-representations as for the $\DZ_6$ group. The fusion rules for surface operators labelled by these representations are given by 
\begin{alignat}{4}
&S_{(\DZ_1,[1])} \times S_{(\DZ_1,[1])}= 6 S_{(\DZ_1,[1])}~,&& ~
S_{(\DZ_1,[1])} \times S_{(\DZ_2,[1])}= 3 S_{(\DZ_1,[1])},~\cr
&S_{(\DZ_1,[1])} \times S_{(\DZ_3,[1])}= 2 S_{(\DZ_1,[1])}~,&& ~
S_{(\DZ_2,[1])} \times S_{(\DZ_2,[1])}= S_{(\DZ_1,[1])} + S_{(\DZ_2,[1])}~,\\
&S_{(\DZ_2,[1])} \times S_{(\DZ_3,[1])}= S_{(\DZ_1,[1])}~,&& ~
S_{(\DZ_3,[1])} \times S_{(\DZ_3,[1])}= 2 S_{(\DZ_3,[1])}~.\nonumber
\end{alignat}
$S_{(S_3,[1])}$ is the trivial surface operator. Note that the non-invertible surface operators $S_{(\DZ_2,[1])}$ and $S_{(\DZ_3,[1])}$ fuse to give a unique outcome. This fusion arises from the semi-direct product structure of the group $S_3 \simeq \DZ_3 \rtimes \DZ_2$. In Section \ref{sec:groupext}, we will show that the fusion rules can be used to determine whether the group is a semi-direct product of smaller groups. 

\subsubsection{$D_8$}

\label{subsec:D8}

$D_8$ has the following normal subgroups
\be
D_4,~ \{[1],[r],[r^2]\}\simeq \DZ_4,~ \{[1],[r^2],[s]\}\simeq \DZ_2 \times \DZ_2,~ \{[1],[r^2],[rs]\} \simeq \DZ_2 \times \DZ_2,~\{1,r^2\}\simeq \DZ_2, \{[1]\} 
\ee
These are fixed under conjugation. $D_8$ also has the non-normal subgroups 
\be 
\{1,s\},\{1,rs\},\{1,r^2s\},\{1,r^3s\}~,
\ee
which are all isomorphic to $\DZ_2$. Under conjugation, these subgroups fall into two classes 
\be
\{\{1,s\},\{1,r^2s\}\}, \{\{1,rs\},\{1,r^3s\}\}~.
\ee
All subgroups have trivial $2^{\text{nd}}$ cohomology group except $H^2(\DZ_2\times \DZ_2,U(1))=\DZ_2$. The irreducible 2-representations are labelled by
\bea
&&(\DZ_1,[1]),~(\DZ_2^{(r^2)},[1]),~(\DZ_2^{(s)},[1]),~(\DZ_2^{(rs)},[1]),~(\DZ_4^{(r)},[1]),~(D_8,[1]),\cr
&&(\DZ_2^{(r^2)}\times \DZ_2^{(s)},[1]),(\DZ_2^{(r^2)}\times \DZ_2^{(s)},[c]),(\DZ_2^{(r^2)}\times \DZ_2^{(rs)},[1]),(\DZ_2^{(r^2)}\times \DZ_2^{(rs)},[d])~.
\eea
where the superscripts of the cyclic components specify the generators of the subgroup. $[c]$ is the non-trivial cohomology class in $H^2(\DZ_2^{(r^2)}\times \DZ_2^{(s)},U(1))$ and  $[d]$ is the non-trivial cohomology class in $H^2(\DZ_2^{(r^2)}\times \DZ_2^{(rs)},U(1))$. The surface operator $S_{(D_8,[1])}$ is the identity under fusion. The fusion of $S_{(\DZ_1,[1])}$ with any surface operator is given by
\be
S_{(\DZ_1,[1])} \times S_{(L,[e])}= \frac{|G|}{|L|} S_{(\DZ_1,[1])}~.
\ee
We also have the fusions
\begin{alignat}{4}
\label{eq:D8fusion}
&S_{(\DZ_2^{(r^2)},[1])} \times S_{(\DZ_2^{(r^2)},[1])}= 4 S_{(\DZ_2^{(r^2)},[1])}~,&&~
S_{(\DZ_2^{(r^2)},[1])} \times S_{(\DZ_2^{(s)},[1])}= 2 S_{(\DZ_1,[1])}~,\cr
&S_{(\DZ_2^{(r^2)},[1])} \times S_{(\DZ_2^{(r)},[1])}= 2 S_{(\DZ_2^{(r^2)},[1])}~,&&~
S_{(\DZ_2^{(r^2)},[1])} \times S_{(\DZ_2^{(r^2)}\times \DZ_2^{(s)},[1])}= 2 S_{(\DZ_2^{(r^2)},[1])}~,\cr
&S_{(\DZ_2^{(r^2)},[1])} \times S_{(\DZ_2^{(r^2)}\times \DZ_2^{(s)},[c])}= 2 S_{(\DZ_2^{(r^2)},[1])}~,&&~
S_{(\DZ_2^{(s)},[1])} \times S_{(\DZ_2^{(s)},[1])}= S_{(\DZ_1,[1])} + 2 S_{(\DZ_2^{(s)},[1])}~,\cr
&S_{(\DZ_2^{(s)},[1])} \times S_{(\DZ_2^{(rs)},[1])}= 2 S_{(\DZ_1,[1])}~,&&~
S_{(\DZ_2^{(s)},[1])} \times S_{(\DZ_4^{(r)},[1])}= S_{(\DZ_1,[1])}~,\cr
&S_{(\DZ_2^{(s)},[1])} \times S_{(\DZ_2^{(r^2)}\times \DZ_2^{(s)},[1])}= 2 S_{(\DZ_2^{(s)},[1])}~,&&~
S_{(\DZ_2^{(s)},[1])} \times S_{(\DZ_2^{(r^2)}\times \DZ_2^{(s)},[c])}= 2 S_{(\DZ_2^{(s)},[1])}~,\\
&S_{(\DZ_2^{(s)},[1])} \times S_{(\DZ_2^{(r^2)}\times \DZ_2^{(rs)},[1])}=  S_{(\DZ_1,[1])}~,&&~
S_{(\DZ_2^{(s)},[1])} \times S_{(\DZ_2^{(r^2)}\times \DZ_2^{(rs)},[d])}=  S_{(\DZ_1,[1])}~,\cr
& S_{(\DZ_2^{(r)},[1])} \times S_{(\DZ_2^{(r)},[1])}=  2 S_{(\DZ_2^{(r)},[1])}~,&&~
S_{(\DZ_2^{(r)},[1])} \times S_{(\DZ_2^{(r^2)}\times \DZ_2^{(s)},[1])}=  S_{(\DZ_2^{(r^2)},[1])}~ , \cr 
&S_{(\DZ_2^{(r^2)}\times \DZ_2^{(s)},[1])} \times S_{(\DZ_2^{(r^2)}\times \DZ_2^{(s)},[1])}=  2 S_{(\DZ_2^{(r^2)}\times \DZ_2^{(s)},[1])}~,&&~
S_{(\DZ_2^{(r)},[1])} \times S_{(\DZ_2^{(r^2)}\times \DZ_2^{(s)},[c])}=  S_{(\DZ_2^{(r^2)},[1])}~,
\cr
& S_{(\DZ_2^{(r^2)}\times \DZ_2^{(s)},[1])} \times S_{(\DZ_2^{(r^2)}\times \DZ_2^{(s)},[c])}=  2 S_{(\DZ_2^{(r^2)}\times \DZ_2^{(s)},[c])}~,&&~
S_{(\DZ_2^{(r^2)}\times \DZ_2^{(s)},[1])} \times S_{(\DZ_2^{(r^2)}\times \DZ_2^{(rs)},[1])}=  S_{(\DZ_2^{(r^2)},[1])}~,\cr
&S_{(\DZ_2^{(r^2)}\times \DZ_2^{(s)},[1])} \times S_{(\DZ_2^{(r^2)}\times \DZ_2^{(rs)},[d])}=  S_{(\DZ_2^{(r^2)},[1])}~,&&~
S_{(\DZ_2^{(r^2)}\times \DZ_2^{(s)},[c])} \times S_{(\DZ_2^{(r^2)}\times \DZ_2^{(rs)},[d])}=  S_{(\DZ_2^{(r^2)},[1])} ~.\nonumber
\end{alignat}
Other fusion rules can be obtained using the permutation $r \to r, s\to rs, c\to d$.
 
\subsubsection{$Q_8$}

$Q_8$ group has the subgroups $\DZ_1,\DZ_2=\{1,ijk\},\DZ_4=\{1,i,i^2,i^3\},\DZ_4=\{1,j,j^2,j^3\},\DZ_4=\{1,k,k^2,k^3\},Q_8$. These are all normal subgroups which are fixed points under conjugation. All of these groups have trivial $2^{\text{nd}}$ cohomology group. Therefore, the irreducible 2-representations are labelled by $(\DZ_1,[1]),(\DZ_2,[1]),(\DZ^{(i)}_4,[1])$,
$(\DZ^{(j)}_4,[1]),(\DZ^{(k)}_4,[1]),(Q_8,[1])$. The fusion rules for surface operators in this case are given by
\begin{alignat}{4}
\label{eq:Q8fusion}
&S_{(\DZ_1,[1])} \times S_{(\DZ_1,[1])}= 8 S_{(\DZ_1,[1])}~,&& ~
S_{(\DZ_1,[1])} \times S_{(\DZ_2,[1])}= 4 S_{(\DZ_1,[1])}~,\cr
&S_{(\DZ_1,[1])} \times S_{(\DZ_4^{(i)},[1])}= 2 S_{(\DZ_1,[1])}~,&& ~
S_{(\DZ_2,[1])} \times S_{(\DZ_2,[1])}=4 S_{(\DZ_2,[1])}~,\\
& S_{(\DZ_2,[1])} \times S_{(\DZ_4^{(i)},[1])}= 2 S_{(\DZ_2,[1])}~,&&~
S_{(\DZ_2,[1])} \times S_{(Q_8,[1])}= S_{(\DZ_2,[1])}~,\cr
& S_{(\DZ_4^{(i)},[1])} \times S_{(\DZ_4^{(i)},[1])}= 2 S_{(\DZ_4^{(i)},[1])}~,&&~
S_{(\DZ_4^{(i)},[1])} \times S_{(\DZ_4^{(j)},[1])}= S_{(\DZ_2,[1])}~.  \nonumber
\end{alignat}
Other fusion rules can be found by permuting $i,j,k$. $S_{(Q_8,[1])}$ is the trivial surface operator. 

We find that the difference in the number of subgroups up to conjugation in $D_8$ and $Q_8$ leads to different numbers of irreducible 2-representations for these groups. Therefore, the fusion of surface operators (even the number of surface operators) can distinguish them. However, as is clear from the examples of $\DZ_6$ and $S_3$, the number of surface operators alone cannot always distinguish two groups. 

\section{$G$ from fusion of surface operators}

\label{sec:GfromSfus}

In this section, we will study surface operator fusion and the properties of the group $G$ that can be reconstructed from it. Let us look at some properties of fusion rules of surface operators. Recall the expression for the fusion of surface operators.
\be
S_{(K,[c])} \times S_{(H,[d])}= \sum_{g\in K\backslash G/H} S_{(K\cap \,\co{g}H,[c \cdot \,\co{g}d])}~.
\ee
From the expression for the fusion coefficient \eqref{eq:fusioncoef}, we find $N_{(K,[c]),(H,[d])}^{(L,[e])} \leq |G|$. When $K=H=\DZ_1$ the 2-cocycle $c$ is trivial, and $|\DZ_1\backslash G/\DZ_1|=|G|$. Therefore, the fusion rule reduces to
\be
S_{(\DZ_1,[1])} \times S_{(\DZ_1,[1])}= |G| ~ S_{(\DZ_1,[1])}~.
\ee
Moreover, $|H\backslash G/K|=|G|$ if and only if $H=K=\DZ_1$. Therefore, $S_{(\DZ_1,[1])}$ is the unique surface operator with the above fusion rule. The order of subgroups of $G$ can be obtained from the fusion rule
\be
\label{eq:ordersub}
S_{(K,[c])} \times S_{(\DZ_1,[1])}= \frac{|G|}{|K|} S_{(\DZ_1,[1])}~.
\ee
Given a 2-representation $(K,[c])$, if the fusion coefficient
\be
\label{eq:trivcoh}
N_{(K,[c]), (K,[c])}^{(K,[c])} \neq 0
\ee
then $[c]=[1]$. Indeed, the non-trivial fusion coefficient above implies that there exists $g \in K\backslash G/K$ such that $[c \cdot \,\co{g}c]=[c]$ which implies that $[c]=[1]$. More generally, given a 2-representation $(K,[c])$, if the fusion coefficient
\be
\label{eq:sameK}
N_{(H,[d]), (K,[c])}^{(H,[d])} \neq 0
\ee
for some $(H,[d])$ such that $|K|=|H|$, then $K$ and $H$ belong to the same conjugacy class of subgroups, and $[c]=[1]$. Indeed, the non-trivial fusion coefficient above implies that there exists $g \in H\backslash G/K$ such that 
\be 
H \cap \,\co{g}K=H, ~ [d \cdot \,\co{g}c]=[d]~.
\ee
Since $|K|=|H|$, $H \cap \,\co{g}K=H$ implies that $\co{g}K=H$. Also, $[d \cdot \,\co{g}c]=[d]$ implies that $[c]=[1]$.
\\

Let us use the properties of the fusion rules of surface operators discussed above to obtain properties of the group $G$ from $N$ condensation surface operators labelled abstractly by $\trep_1,...,\trep_N$ and their fusions
\be
S_{\trep_1} \timess S_{\trep_2} = \sum_{i=1}^N N_{\trep_1,\trep_2}^{\trep_i} ~ S_{\trep_i} ~.
\ee
A 2-representation $\trep$ corresponds to some $K_{\trep} \subseteq G$ and $[c_{\trep}]\in H^2(K_{\trep},U(1))$. First of all, there is a unique surface operator which is the identity under fusion. We will label the corresponding 2-representation $\mathds{1}$. We know that 
\be
K_{\mathds{1}}=G, ~ [c_{\mathds{1}}]=[1]~.
\ee
Also, there is a unique surface operator, say $S_{\trmax}$ such that the fusion coefficient $N_{\trmax,\trmax}^{\trmax}$ takes its maximal value. Then,
\be
|G|= N_{\trmax, \trmax}^{\trmax}~.
\ee
We know that 
\be
K_{\trmax}=\DZ_1, ~ [c_{\trmax}]=[1]~.
\ee
By identifying the surface operator $S_{\trmax}$ among all the surface operators, we can obtain the order of $K_{\trep}$ for any 2-representation $\trep$ using 
\be
\label{eq:subgsize}
|K_{\trep}|= \frac{|G|}{N_{\trep, \trmax}^{\trmax}}~.
\ee
Also, from the discussion around \eqref{eq:trivcoh}, we have that $[c_{\trep}]=[1]$ if any only if 
\be
\label{eq:samesubg}
N_{\trep, \trep}^{\trep} \neq 0~.
\ee
Since we can identify the subset of 2-representations with trivial $[c_{\trep}]$, we can construct the Burnside ring of $G$ from surface operator fusion. 

 For two 2-representations $\trep$ and $\trep'$ such that $|K_{\trep}|=|K_{\trep}|$, $K_{\trep}$ and $K_{\trep'}$ belongs to the same conjugacy class of subgroups if and only if 
\be
N_{\trep, \trep'}^\trep \neq 0~.
\ee
This criterion allows us to find all 2-representations corresponding to the same $K_{\trep}$ up to conjugation.

Can we determine whether the group $G$ is abelian? In the following subsections, we will show that all normal subgroups of $G$ and their intersections can be identified from the fusion rules of surface operators. This information can then be used to determine whether $G$ is abelian. If $G$ is abelian, then we will show that $G$ can be constructed from the fusion rules up to isomorphism. 

\subsection{Normal Subgroups from Fusion Rules}

\label{sec:normalfrfusion}

We can infer the existence or lack of normal subgroups from the fusion rules of surface operators. 

\textit{A surface operator $S_{\trep}$ satisfies the fusion rule
\be
\label{eg:normalfus}
S_{\trep} \timess  S_{\trep} = \sum_{i=1}^{\frac{|G|}{|K_{\trep}|}} S_{\trep_i}~,
\ee
where $K_{\trep_i}$ and $K_{\trep}$ belong to the same conjugacy class of subgroups for all $i$ if and only if $K_{\trep}$ is a normal subgroup of $G$.
}

\noindent To see this, consider the fusion rule
\be
S_{(K_{\trep},[c_{\trep}])} \timess  S_{(K_{\trep},[c_{\trep}])} = \sum_{i=1}^{\frac{|G|}{|K_{\trep}|}} S_{(K_{\trep},[c_{\trep_i}])}~.
\ee
From the expression for the fusion rules \eqref{eq:surfacefusion}, we get
\be
\co{g}K_{\trep}=K_{\trep} ~ \forall ~ g \in K_{\trep}\backslash G / K_{\trep} \implies \co{g}K_{\trep}=K_{\trep} ~ \forall ~ g \in G~,
\ee
which implies that $K_{\trep}$ is a normal subgroup. Conversely, let us consider a normal subgroup $K_{\trep} \subset G$. Using \eqref{eq:surfacefusion}, we get the fusion rules 
\be
S_{(K_{\trep},[c_{\trep}])} \timess  S_{(K_{\trep},[c_{\trep}])} = \sum_{i=1}^{\frac{|G|}{|K_{\trep}|}} S_{(K_{\trep},[c_{\trep_i}])}~.
\ee

If $N_{\trep,\trep'}^{\trep''}\neq 0$ where the 2-representations $\trep$ and $\trep'$ corresponds to normal subgroups $K_{\trep},K_{\trep'}\subseteq G$, then the expression for the fusion rule \eqref{eq:surfacefusion} implies 
\be
K_{\trep''}= K_{\trep} \cap K_{\trep'}~.
\ee
This shows that the modular lattice of normal subgroups of $G$ can be determined from the fusion rules of surface operators.  Recall that the modular lattice of normal subgroups of $G$ can also be determined from the character table (see section \ref{sec:Wilson}). More generally, if $K_{\trep}$ is not normal and $K_{\trep'}$ is normal, then we get
\be
K_{\trep''}= K_{\trep} \cap K_{\trep'}~,
\ee
and then
\be
\label{intsize}
|K_{\trep} \cap K_{\trep'}|= \frac{|G|}{N_{\trep'',\trmax}^{\trmax}}
\ee
for some $\trep$. Therefore, the order of intersection of a normal subgroup with a non-normal subgroup can be determined from the fusion rules of surface operators. Note that this information cannot be obtained from the character table (see the $D_8$ v/s $Q_8$ discussion in section \ref{sec:Wilson}).

\subsection{2-representations of $G/N$}
\label{sec:quotientfus}

In the previous subsection, we learned that given the fusion rules surface operators $S_{\trep}$ we can identify the normal subgroups of $G$. In this section, we will show that given a normal subgroup $N$ of $G$, the irreducible 2-representations of the quotient group $G/N$ are also determined by the 2-representations of $G$. Irreducible 2-representations of $G/N$ are labelled by $(M,[d])$ where $M$ is a subgroup of $G$ and $d \in H^2(M,U(1))$. The correspondence theorem states that there is a bijection between the subgroups of $G$ containing $N$ and subgroups of $G/N$ (for example see \cite{humphreys1996course}). For every subgroup $M \subseteq G/N$ we have some $K \subseteq G$ containing $N$ such that
\be
K/N \cong M~.
\ee
This is a bijection. Therefore, the irreducible 2-representations of $G/N$ can be equivalently labelled as $(K/N,[r])$ where $[r]\in H^2(K/N,U(1))$. A 2-representation $(K/N,[r])$ of $G/N$ corresponds to a 2-representation $(K,[c])$ of $G$ where $[c] \in H^2(K,U(1))$ is the cohomology class of the inflation of the 2-cocycle $r$ given by
\be
c = r \circ \pi~,
\ee
where $\pi: G \to G/N$ is the projection $\pi(g)=Ng$. If there exists $\pi(l)\in G/N$ such that 
\be
(K/N,[r]) = (\co{\pi(l)}(H/N),[\co{\pi(l)}p])~.
\ee
then we have the relation 
\be
\pi(l)\pi(h)\pi(l)^{-1}=\pi(lhl^{-1})=\pi(k)
\ee
for $h\in H$ and $k \in K$. This implies that $lhl^{-1}=n k$ for some $n \in N$. Since $N \subseteq K$, $k':=nk\in K$. Therefore, $lhl^{-1}=k'$ which shows that the subgroups $H$ and $K$ of $G$ are conjugate. Also, the inflation of 2-cocycles $r$ and $p$ to $H^2(K,U(1))$ and $H^2(H,U(1))$, respectively are mapped to each other under the conjugation $\co{l}H=K$. This shows that the equivalent 2-representations $(K/N,[r]) $ and $(H/N,[p])$  of $G/N$ correspond to equivalent 2-representations of $(K,[c])$ and $(H,[d])$ of $G$, respectively, where $c:=r \circ \pi$ and $d:= p \circ \pi$.  

Consider the fusion coefficient 
\be
N_{(K/N,[r]),(H/N,[s])}^{(L/N,[p])}=|\{l \in (K/N)\backslash (G/N) /(H/N) ~|~ (K/N) \cap \,\co{l}(H/N)=(L/N), [r \cdot \,\co{l}s]=[r]\}|
\ee
for three 2-representations $(K/N,[r]),(H/N,[s])$ and $(L/N,[r])$ of $G/N$. Let $[c],[d],[e]$ be the inflation of the 2-cocycles $[r],[s],[p]$, respectively. We have
\be
N_{(K,[c]),(H,[d])}^{(L,[e])}=|\{g \in K\backslash G /H ~|~ K \cap \,\co{g}H=L, [c \cdot \,\co{g}d]=[e]|~.
\ee
The double coset $K\backslash G/H$ is the set of equivalence classes under the equivalence relation $g_1 \sim g_2$ if $g_1=kg_2h$ for some $k\in K,h\in H $. If $N$ is a normal subgroup of $K,H$ and $G$, then
\be 
kg_1h=g_2 \implies \pi(kg_1h)=\pi(g_2) \implies \pi(k)\pi(g_1)\pi(h)=\pi(g_2) 
\ee
Note that $\pi(k)\in K/N, \pi(h)\in H/N$ and $\pi(g_1),\pi(g_2)\in G/N$. Therefore, if $g_1 \sim g_2$ then $\pi(g_1)$ and $\pi(g_2)$ are in the same equivalence class in $(K/N) \backslash (G/N)/(H/N)$. Similarly, if $g_1 \not\sim g_2$ then $\pi(g_1) \not\sim \pi(g_2)$. The projection $\pi$ defines a bijective map between the sets $K\backslash G/H$ and $(K/N)\backslash (G/N) /(H/N)$.
Now, if $g \in K\backslash G /H$ satisfies
\be 
K \cap H^g=L, [c \cdot d^g]=[e]~,
\ee
then
\bea
K \cap \,\co{g}H=L \Leftrightarrow (K \cap \,\co{g}H)/N=L/N  &\Leftrightarrow& (K/N) \cap (\co{g}H)/N=L/N \cr &\Leftrightarrow& (K/N) \cap \,\co{\pi(g)}(H/N)=L/N 
\eea
and
\bea
[c \cdot \,\co{g}d]=[e] \Leftrightarrow  [r \circ \pi \cdot \,\co{g}(s \circ \pi)]=[p \circ \pi] \Leftrightarrow [r \cdot \,\co{\pi(g)}s]=[p]~.
\eea
where $\pi(g)\in (K/N)\backslash (G/N) /(H/N)$. Therefore, 
\be
N_{(K/N,[r]),(H/N,[s])}^{(L/N,[p])}= N_{(K,[c]),(H,[d])}^{(L,[e])}~.
\ee
This shows that the subring 2Rep$(G/N)$ is contained in 2Rep$(G)$ for any normal subgroup $N  \vartriangleleft G$. Given a 2-representation $\trep$, how do we determine if $\trep\in \text{2Rep}(G/N)$? First of all, if $\trep\in \text{2Rep}(G/N)$ then $K_{\trep}$ should contain the normal subgroup $N$. This can be checked using the fusion rules of surface operators and equation \eqref{intsize}. Finally, we also require $[c_{\trep}]$ to be an inflation of a 2-cocycle in $H^2(K_{\trep}/N,U(1))$. A necessary condition for this is that the restriction of the 2-cocycle $c_\trep$ to the normal subgroup $N$ should be trivial. This can be checked using the fusion rule
\be
S_{\trep} \times S_{\trep_N}= S_{\trep'}~,
\ee
where $\trep_N$ is the 2-representation satisfying $K_{\trep_N}=N$ and $[c_{\trep_N}]=[1]$. Then, using \eqref{eq:surfacefusion} we find that $[c_{\trep'}]=[1]$ if and only if $[c_\trep|_{N}]=[1]$.

Note that the 2-representations of $N$ itself cannot be determined in general. This is because the 2-representations of $G$ only ``knows" about subgroups of $N$ up to conjugation by $G$. However, the 2-representations of $N$ are determined by subgroups of $N$ up to conjugation by elements of $N$. In general, the latter set is larger than the former.

\subsection{Group Extensions and Fusion Rules}

\label{sec:groupext}

If a group $G$ can be obtained from smaller groups through group extensions (including direct and semi-direct products), then $G$ contains normal subgroups. Now that we know how to determine the existence or lack of normal subgroups from fusion rules, we can consider the following cases. Suppose there are no fusion rules of the form \eqref{eg:normalfus} except for $K_{\trep}=G,\DZ_1$. This implies that the group $G$ does not contain any proper normal subgroups. Therefore, $G$ is a simple group and cannot be written as a group extension of smaller groups. In the following subsections, we will consider the case where one or more fusions of the form \eqref{eg:normalfus} do exist for a proper non-trivial subgroup $N \subset G$. Then, we will show that similar fusion rules can be used to determine whether the group $G$ can be written as a direct product, semi-direct product or a non-split group extension. 

\subsubsection{Direct Product}

Suppose the group $G$ is a direct product of groups $N,K$. Then we have two normal subgroups in $G$ isomorphic to $N$ and $K$. We will abuse notation and label these normal subgroups by $N,K$ itself. Moreover, these normal subgroups should satisfy 
\be
N \cap K= \DZ_1, ~ G=NK ~.
\ee
The second condition is equivalent to requiring $|G|=|N||K|$ (Recall that $|NH|=\frac{|N||H|}{|N\cap H|}$). Therefore, the group $G$ is a direct product of $N,K$ if and only if the following three conditions are satisfied 
\begin{enumerate}
\item $N \triangleleft G,K \triangleleft G$~.
\item $N \cap K = \DZ_1$~.
\item $|G|=|N||K|$~.
\end{enumerate}
The direct product structure of a group can be determined from the fusion of surface operators. 

\textit{Two 2-representations $\trep$ and $\trep'$ satisfy the fusion rules
\bea
\label{eq:dpfusions}
S_{\trep} \timess  S_{\trep} = \sum_{i=1}^{\frac{|G|}{|K_{\trep}|}} S_{\trep_i} ~,\cr
S_{\trep'} \timess  S_{\trep'} = \sum_{j=1}^{\frac{|G|}{|K_{\trep'}|}} S_{\trep'_j}~, \\
S_{\trep} \timess  S_{\trep'} = S_{\trmax}~. \nonumber
\eea
where $K_{\trep_i}$ and $K_{\trep}$ belong to the same conjugacy class of subgroups for all $i$, and $K_{\trep'_j}$ and $K_{\trep'}$ belong to the same conjugacy class of subgroups for all $j$ if and only if $G=K_{\trep} \times K_{\trep'}$.}

To understand this statement, assume the three types of fusions above. From section \ref{sec:normalfrfusion} we know that first and second fusion implies that $K_{\trep}$ and $K_{\trep'}$ are normal subgroups of $G$. The third fusion above implies that (using \eqref{eq:surfacefusion})
\be
|K_{\trep}\backslash G/ K_{\trep'}|=1 \text{ and } K_{\trep} \cap K_{\trep'}=\DZ_1~.
\ee
Since $K_\trep$ and $K_{\trep'}$ are normal subgroups, we have
\be
|K_{\trep}\backslash G/ K_{\trep'}|= \frac{|G|}{|K_{\trep}K_{\trep'}|}~.
\ee
The fusion rules above imply that $|K_\trep K_{\trep'}|=|G|$ which using $K_{\trep} \cap K_{\trep'}=\DZ_1$ implies $|K_{\trep}| |K_{\trep'}|=|G|$. Therefore, the normal subgroups $K_\trep$ and $K_{\trep'}$ satisfies the three conditions for $G$ to be a direct product $K_\trep \times K_{\trep'}$.

Conversely, if $G= K_{\trep} \times K_{\trep'}$, then $K_{\trep}$ and $K_{\trep'}$ are normal subgroups satisfying $K_{\trep} \cap K_{\trep'}=\DZ_1$ and $|K_\trep||K_{\trep'}|=|G|$. Then, using \eqref{eq:surfacefusion} it is easily verified that the corresponding 2-representations $\trep$ and $\trep'$ satisfy the fusion rules \eqref{eq:dpfusions}.
\\

\noindent {\bf Reconstructing abelian $G$}
\vspace{0.2cm}

From section \ref{sec:normalfrfusion}, we know that the order of all normal subgroups of $G$ and their intersections can be determined from the fusion rules of surface operators. In particular, we can check whether all subgroups are normal. If not, then the group $G$ is not abelian. If all subgroups are normal, then $G$ is of the form $Q_8 \times A$ where $A$ is an abelian group\footnote{The abelian group $A$ has some additional properties but we will not need it in our discussion.} \cite{dedekind1897ueber}. If $G=Q_8 \times A$, then $Q_8$ and all subgroups of $Q_8$ are normal in $G$. In particular, the fusion rules of surface operators labelled by 2-representations in 2Rep$(G)$ contain the fusion rules \eqref{eq:Q8fusion}. Let us show that $Q_8$ is the unique group of order $8$ with the set of surface operators and fusion rules given in \eqref{eq:Q8fusion}. Indeed, the groups of order 8 up to isomorphism are $\DZ_8, \DZ_2 \times \DZ_4, \DZ_2 \times \DZ_2 \times \DZ_2, D_8$ and $Q_8$. The groups $\DZ_8, \DZ_4 \times \DZ_2$ and $\DZ_2 \times \DZ_2 \times \DZ_2$ have $4,8$ and $16$ subgroups, respectively. Therefore, these groups have $4,8$ and $16$ irreducible 2-representations, respectively. Also, from the discussion in section \ref{subsec:D8} we know that $D_8$ has $10$ irreducible 2-representations. From \eqref{eq:Q8fusion}, we know that $Q_8$ has 6 irreducible 2-representations. This implies that $Q_8$ can be distinguished from all other groups of order $8$ by just counting the number of 2-representations. Therefore, we can use the fusion rules of surface operators to determine whether $Q_8$ is a subgroup of $G$, which in turn determines whether $G$ is abelian. 
 
If $G$ is abelian, then we can determine the group $G$ up to isomorphism as follows. $G$ is a direct product of the form
\be
G=\DZ_{p_1^{n_1}} \times \DZ_{p_2^{n_2}}\times ...\times \DZ_{p_k^{n_k}}~,
\ee
where $|G|=p_1^{n_1}p_2^{n_2}...p_k^{n_k}$ is a decomposition of the order of $G$ into prime powers. The order of normal subgroups and their intersections can be found from the fusion rules using the discussion in section \ref{sec:normalfrfusion}. Therefore, the integers $p_i$ and $n_i$ can be determined from the fusion rules of surface operators. Note that we only need the fusion rules of surface operators $S_{\trep}$ with $[c_{\trep}]=[1]$ to determine the abelian group $G$. In other words, the Burnside ring of $G$ determines an abelian group. This is in agreement with \cite{raggi2005groups}, where the author shows that the Burnside ring determines a group $G$ if $G$ is abelian, Hamiltonian or a minimal simple group.  
\\

\noindent {\bf Derived Subgroup}
\vspace{0.2cm}

Now that we know how to determine whether a group is abelian or not using its 2-representations, we can also determine the derived subgroup of $G$ from 2-representations. The derived subgroup $[G,G]$ is the smallest normal subgroup such that $G/[G,G]$ is abelian. Using the results in section \ref{sec:normalfrfusion}, we can identify all normal subgroups of $G$ and their orders from the fusion rules. Also, using the results in section \ref{sec:quotientfus}, we can also construct all 2-representations of the quotient group and their fusion rules. This then allows us to determine whether the quotient group is abelian. Therefore, the derived subgroup can be determined from the fusion rules of 2-representations of $G$.

\subsubsection{Semi-Direct Product and Group Extensions}

Suppose the group $G$ is a semi-direct product of groups $N,K$. Then we have two subgroups in $G$ isomorphic to $N$ and $K$. We will abuse notation and label these subgroups by $N,K$ itself. Moreover, these subgroups should satisfy 
\be
N \triangleleft G,~ N \cap K= 1, ~ G=NK ~.
\ee
Note that the definition is the same as in the case of the direct product except that $K$ is not normal. From \eqref{eq:dpfusions} it follows that we have the fusions 
\bea
\label{eq:sdpfusions}
S_{\trep} \timess  S_{\trep} = \sum_{i=1}^{\frac{|G|}{|K_{\trep}|}} S_{\trep_i} ~,\cr 
S_{\trep} \timess  S_{\trep'} = S_{\trmax}~. 
\eea
where $K_{\trep_i}$ and $K_{\trep}$ belong to the same conjugacy class of subgroups for all $i$ if and only if $G=K_{\trep} \rtimes K_{\trep'}$. Note that since $K_{\trep'}$ is not normal, it doesn't satisfy fusion rules of the form \eqref{eg:normalfus}. 

The most general case is when $G$ is a non-split extension of groups $N,K$. $G$ has a normal subgroup isomorphic to $N$ which we will denote as $N$ itself and $G/N \simeq K$. In this case, $K$ is not a subgroup of $G$. If fusion rules of surface operators do not contain fusions of the form \eqref{eq:dpfusions} or \eqref{eq:sdpfusions} then the gauge group $G$ does not admit a presentation as a direct or semi-direct product. A summary of results in this section is given in fig. \ref{fig:ressummary}.

\begin{figure}
    \centering

\tikzset{every picture/.style={line width=0.75pt}} 

\begin{tikzpicture}[x=0.75pt,y=0.75pt,yscale=-0.9,xscale=1]

\draw  [color={rgb, 255:red, 245; green, 166; blue, 35 }  ,draw opacity=1 ][fill={rgb, 255:red, 248; green, 198; blue, 155 }  ,fill opacity=1 ] (60.99,144.54) -- (61,97.77) -- (111,75.44) -- (110.99,122.21) -- cycle ;
\draw  [color={rgb, 255:red, 245; green, 166; blue, 35 }  ,draw opacity=1 ][fill={rgb, 255:red, 248; green, 198; blue, 155 }  ,fill opacity=1 ] (101.55,148.51) -- (101.56,101.74) -- (151.55,79.41) -- (151.54,126.18) -- cycle ;
\draw  [color={rgb, 255:red, 245; green, 166; blue, 35 }  ,draw opacity=1 ][fill={rgb, 255:red, 248; green, 198; blue, 155 }  ,fill opacity=1 ] (254,149) -- (254.01,102.23) -- (304,79.9) -- (303.99,126.67) -- cycle ;
\draw    (381,11) -- (380,842) ;
\draw    (19,49) -- (640,50) ;
\draw  [color={rgb, 255:red, 245; green, 166; blue, 35 }  ,draw opacity=1 ][fill={rgb, 255:red, 248; green, 198; blue, 155 }  ,fill opacity=1 ] (60.99,240.54) -- (61,193.77) -- (111,171.44) -- (110.99,218.21) -- cycle ;
\draw  [color={rgb, 255:red, 245; green, 166; blue, 35 }  ,draw opacity=1 ][fill={rgb, 255:red, 248; green, 198; blue, 155 }  ,fill opacity=1 ] (101.55,244.51) -- (101.56,197.74) -- (151.55,175.41) -- (151.54,222.18) -- cycle ;
\draw  [color={rgb, 255:red, 245; green, 166; blue, 35 }  ,draw opacity=1 ][fill={rgb, 255:red, 248; green, 198; blue, 155 }  ,fill opacity=1 ] (254,245) -- (254.01,198.23) -- (304,175.9) -- (303.99,222.67) -- cycle ;
\draw  [color={rgb, 255:red, 245; green, 166; blue, 35 }  ,draw opacity=1 ][fill={rgb, 255:red, 248; green, 198; blue, 155 }  ,fill opacity=1 ] (90.99,342.54) -- (91,295.77) -- (141,273.44) -- (140.99,320.21) -- cycle ;
\draw  [color={rgb, 255:red, 245; green, 166; blue, 35 }  ,draw opacity=1 ][fill={rgb, 255:red, 248; green, 198; blue, 155 }  ,fill opacity=1 ] (131.55,346.51) -- (131.56,299.74) -- (181.55,277.41) -- (181.54,324.18) -- cycle ;
\draw  [color={rgb, 255:red, 245; green, 166; blue, 35 }  ,draw opacity=1 ][fill={rgb, 255:red, 248; green, 198; blue, 155 }  ,fill opacity=1 ] (224,344) -- (224.01,297.23) -- (274,274.9) -- (273.99,321.67) -- cycle ;
\draw  [color={rgb, 255:red, 245; green, 166; blue, 35 }  ,draw opacity=1 ][fill={rgb, 255:red, 248; green, 198; blue, 155 }  ,fill opacity=1 ] (58.99,441.54) -- (59,394.77) -- (109,372.44) -- (108.99,419.21) -- cycle ;
\draw  [color={rgb, 255:red, 245; green, 166; blue, 35 }  ,draw opacity=1 ][fill={rgb, 255:red, 248; green, 198; blue, 155 }  ,fill opacity=1 ] (99.55,445.51) -- (99.56,398.74) -- (149.55,376.41) -- (149.54,423.18) -- cycle ;
\draw  [color={rgb, 255:red, 245; green, 166; blue, 35 }  ,draw opacity=1 ][fill={rgb, 255:red, 248; green, 198; blue, 155 }  ,fill opacity=1 ] (175,443) -- (175.01,396.23) -- (225,373.9) -- (224.99,420.67) -- cycle ;
\draw  [color={rgb, 255:red, 245; green, 166; blue, 35 }  ,draw opacity=1 ][fill={rgb, 255:red, 248; green, 198; blue, 155 }  ,fill opacity=1 ] (55.99,544.54) -- (56,497.77) -- (106,475.44) -- (105.99,522.21) -- cycle ;
\draw  [color={rgb, 255:red, 245; green, 166; blue, 35 }  ,draw opacity=1 ][fill={rgb, 255:red, 248; green, 198; blue, 155 }  ,fill opacity=1 ] (96.55,548.51) -- (96.56,501.74) -- (146.55,479.41) -- (146.54,526.18) -- cycle ;
\draw  [color={rgb, 255:red, 245; green, 166; blue, 35 }  ,draw opacity=1 ][fill={rgb, 255:red, 248; green, 198; blue, 155 }  ,fill opacity=1 ] (247,547) -- (247.01,500.23) -- (297,477.9) -- (296.99,524.67) -- cycle ;
\draw  [color={rgb, 255:red, 245; green, 166; blue, 35 }  ,draw opacity=1 ][fill={rgb, 255:red, 248; green, 198; blue, 155 }  ,fill opacity=1 ] (63.99,681.54) -- (64,634.77) -- (114,612.44) -- (113.99,659.21) -- cycle ;
\draw  [color={rgb, 255:red, 245; green, 166; blue, 35 }  ,draw opacity=1 ][fill={rgb, 255:red, 248; green, 198; blue, 155 }  ,fill opacity=1 ] (104.55,685.51) -- (104.56,638.74) -- (154.55,616.41) -- (154.54,663.18) -- cycle ;
\draw  [color={rgb, 255:red, 245; green, 166; blue, 35 }  ,draw opacity=1 ][fill={rgb, 255:red, 248; green, 198; blue, 155 }  ,fill opacity=1 ] (229,679) -- (229.01,632.23) -- (279,609.9) -- (278.99,656.67) -- cycle ;

\draw (159.11,103.82) node [anchor=north west][inner sep=0.75pt]    {$=$};
\draw (181.33,102.05) node [anchor=north west][inner sep=0.75pt]    {$N_{\rho _{\mathrm{max}} ,\rho _{\mathrm{max}}}^{\rho _{\mathrm{max}}}$};
\draw (104.56,97.83) node [anchor=north west][inner sep=0.75pt]    {$S_{\rho _{\mathrm{max}}{}}$};
\draw (124,25.4) node [anchor=north west][inner sep=0.75pt]    {Type of Fusion};
\draw (475,23.4) node [anchor=north west][inner sep=0.75pt]    {Property of $G$};
\draw (473,100.4) node [anchor=north west][inner sep=0.75pt]    {$|G|=N_{\rho _{\mathrm{max}} ,\rho _{\mathrm{max}}}^{\rho _{\mathrm{max}}}$};
\draw (63,192.86) node [anchor=north west][inner sep=0.75pt]    {$S_{\rho }$};
\draw (158.11,197.82) node [anchor=north west][inner sep=0.75pt]    {$=$};
\draw (104.56,195.83) node [anchor=north west][inner sep=0.75pt]    {$S_{\rho _{\mathrm{max}}{}{}}$};
\draw (256.56,195.32) node [anchor=north west][inner sep=0.75pt]    {$S_{\rho _{\mathrm{max}}{}}$};
\draw (63,96.17) node [anchor=north west][inner sep=0.75pt]    {$S_{\rho _{\mathrm{max}}{}}$};
\draw (256.01,100.63) node [anchor=north west][inner sep=0.75pt]    {$S_{\rho _{\mathrm{max}}{}}$};
\draw (179.33,195.05) node [anchor=north west][inner sep=0.75pt]    {$N_{\rho ,\rho _{\mathrm{max}}}^{\rho _{\mathrm{max}}}$};
\draw (467,188.4) node [anchor=north west][inner sep=0.75pt]    {$|K_{\rho } |=\frac{|G|}{N_{\rho ,\rho _{\mathrm{max}}}^{\rho _{\mathrm{max}}}}$};
\draw (93,293.86) node [anchor=north west][inner sep=0.75pt]    {$S_{\rho }$};
\draw (195.11,298.82) node [anchor=north west][inner sep=0.75pt]    {$\ni $};
\draw (134.56,296.83) node [anchor=north west][inner sep=0.75pt]    {$S_{\rho {}{}}$};
\draw (226.01,295.63) node [anchor=north west][inner sep=0.75pt]    {$S_{\rho {}{}}$};
\draw (469,306.4) node [anchor=north west][inner sep=0.75pt]    {$\rho =( K_{\rho } ,[ 1])$};
\draw (10,103.4) node [anchor=north west][inner sep=0.75pt]    {$1.$};
\draw (10,202.4) node [anchor=north west][inner sep=0.75pt]    {$2.$};
\draw (10,303.4) node [anchor=north west][inner sep=0.75pt]    {$3.$};
\draw (10,406.4) node [anchor=north west][inner sep=0.75pt]    {$4.$};
\draw (61,392.86) node [anchor=north west][inner sep=0.75pt]    {$S_{\rho }$};
\draw (155.11,397.82) node [anchor=north west][inner sep=0.75pt]    {$\ni $};
\draw (102.56,395.83) node [anchor=north west][inner sep=0.75pt]    {$S_{\rho '{}{}}$};
\draw (177.01,399.63) node [anchor=north west][inner sep=0.75pt]    {$S_{\rho {}{}}$};
\draw (232,400.4) node [anchor=north west][inner sep=0.75pt]    {\hspace{0.15cm} and $|K_{\rho } |=|K_{\rho'}|~.$};
\draw (475,404.4) node [anchor=north west][inner sep=0.75pt]    {$K_{\rho } \ \sim \ K_{\rho '}$};
\draw (58,496.86) node [anchor=north west][inner sep=0.75pt]    {$S_{\rho }$};
\draw (162.11,513.82) node [anchor=north west][inner sep=0.75pt]    {$=$};
\draw (99.56,499.83) node [anchor=north west][inner sep=0.75pt]    {$S_{\rho {}{}}$};
\draw (249.56,497.32) node [anchor=north west][inner sep=0.75pt]    {$S_{\rho _{i}{}}$};
\draw (9,514.4) node [anchor=north west][inner sep=0.75pt]    {$5.$};
\draw (185,494.4) node [anchor=north west][inner sep=0.75pt]    {$\sum {_{i=1}^{\frac{|G|}{|K_{\rho } |}}}$};
\draw (103,569.4) node [anchor=north west][inner sep=0.75pt]    {\hspace{0.2cm} and $K_{\rho } \sim K_{\rho _{i}} \ \forall \ i~.$};
\draw (445,516.4) node [anchor=north west][inner sep=0.75pt]    {$K_{\rho}$ is a normal subgroup.};
\draw (66,633.86) node [anchor=north west][inner sep=0.75pt]    {$S_{\rho }$};
\draw (181.11,638.82) node [anchor=north west][inner sep=0.75pt]    {$=$};
\draw (107.56,636.83) node [anchor=north west][inner sep=0.75pt]    {$S_{\rho '{}}$};
\draw (231.56,629.32) node [anchor=north west][inner sep=0.75pt]    {$S_{\rho _{\mathrm{max}}{}}$};
\draw (12,643.4) node [anchor=north west][inner sep=0.75pt]    {$6.$};
\draw (57,714.4) node [anchor=north west][inner sep=0.75pt]    {$6.1\ \ \ \ and\ \ S_{\rho } ,S_{\rho } '$ satisfies fusion type $5~.$};
\draw (466,715.4) node [anchor=north west][inner sep=0.75pt]    {$G=K_{\rho } \times K_{\rho '}$};
\draw (57,755.4) node [anchor=north west][inner sep=0.75pt]    {$6.2\ \ \ \ and\ \ S_{\rho }$ satisfies fusion type $5~.$};
\draw (464,756.4) node [anchor=north west][inner sep=0.75pt]    {$G=K_{\rho } \rtimes K_{\rho '}$};
\draw (56,794.4) node [anchor=north west][inner sep=0.75pt]    {$6.3$ \hspace{0.3cm} No $S_{\rho}$  satisfying fusion type $5$ and $6~.$};
\draw (413,795.4) node [anchor=north west][inner sep=0.75pt]    {$G$ is a non-split group extension.};

\end{tikzpicture}
    \caption{Properties of $G$ from fusion rules of surface operators. Notations used are $\sim$ : equality up to conjugation, $\ni$ : contained in the fusion outcome.}
    \label{fig:ressummary}
\end{figure}

\subsection{Distinguishing groups with isomorphic Wilson line fusion}

\subsubsection{$D_n$ v/s $Q_n$}

As we saw in section \ref{sec:examples}, the $D_8$ and $Q_8$ groups can be distinguished using the number of 2-representations. To relate this to our discussion in section \ref{sec:groupext} note that $D_8 \simeq \DZ_4 \rtimes \DZ_2$ while $Q_8$ is not a semi-direct product. Since fusion rules of surface operators can be used to distinguish a group admitting a semi-direct product presentation from a group that is not a semi-direct product, it follows that $D_8$ and $Q_8$ can be distinguished. 

More generally, let us consider the Dihedral groups $D_n$ and generalized quaternion group $Q_m$. These groups have the presentations
\bea
D_n:= \langle r,s | r^{\frac{n}{2}}=s^2=(sr)^2=1\rangle~, \\
Q_{2^m}:= \langle a,b| a^{2^{n-1}}=b^4=1, a^{2^{n-2}}=b^2, bab^{-1}=a^{-1} \rangle ~,
\eea
where $n$ is an even integer and $m$ is an arbitrary integer. It is known that  $D_{2^r}$ and $Q_{2^r}$ for all $r\geq 3$ are non-isomorphic and have the same character table (see page 64 of \cite{feit1965characters}). 
However, we can use the fusion of surface operators to distinguish this infinite family of pairs of groups. This is because the dihedral group $D_n$ is a semi-direct product $\DZ_n \rtimes \DZ_2$ while the generalized quaternion group cannot be written as a semi-direct product. To understand this statement, note that any $x\in Q_{2^r}$ has order $2^l$ for some $1\leq l \leq 2^r$. Then the order of $x^{2^{l-1}}$ is $2$. It follows that any subgroup of $Q_{2^r}$ contains order two elements. If $Q_{2^r}=N \rtimes K$ for some subgroups $N$ and $K$ and $N \cap K=1$, then $Q_{2^r}$ should have at least two order two elements. However, $Q_{2^r}$ has only one element of order two. Therefore, $Q_{2^r}$ is not a semi-direct product. 
 
\subsubsection{Groups of order $p^3$}

Note that both $D_8$ and $Q_8$ are groups of order $2^3$. We can generalize this example to study groups of order $p^3$ for any prime $p$. It is known that there are only two non-abelian groups of order $p^3$ \cite{conrad2014groups}. Both of them have the same character table (see page 562, 16.14 in \cite{huppert2013endliche}) and hence cannot be distinguished using the fusion of Wilson lines. Both are semi-direct products as well. One is the Heisenberg group $H_p:=(\DZ_p \times \DZ_p) \rtimes \DZ_p$ while the other one is $G_p:= \DZ_{p^2} \rtimes \DZ_{p}$. $H_p$ and $G_p$ can be distinguished using the fact that all non-trivial elements of $H_p$ have order $p$ while $G_p$ has a normal subgroup isomorphic to $\DZ_{p^2}$. 

Suppose we are given a set of fusion rules for surface operators $S_{\trep}$ labelled by 2-representations in 2Rep$(G)$ for some non-abelian group of order $p^3$. Using \eqref{eq:subgsize} we can determine the surface operators for which $K_{\trep}$ has order $p^2$. $K_{\trep}$ is isomorphic to either $\DZ_{p^2}$ or $\DZ_p \times \DZ_p$. Also, $H^2(\DZ_{p^2},U(1))=\DZ_1$ while $H^2(\DZ_p\times \DZ_p,U(1))=\DZ_{p}$. Using \eqref{eq:samesubg}, we can determine distinct 2-representations with the same $K_{\trep}$ but different $c_{\rho}$. If we find only one 2-representation with the same $K_{\trep}$, then this implies that $H^2(K_{\trep},U(1))=\DZ_{1}$ and therefore $K_{\trep}=\DZ_{p^2}$ which in turn implies that the group $G=G_p$. Otherwise, $G=H_p$.

\subsubsection{Isocategorical Groups}

Two groups $G$ and $G'$ are called isocategorical if they have the same tensor category of representations. Such groups were defined and characterised in \cite{etingof2001isocategorical}. They find an infinite family of non-isomorphic isocategorical groups constructed as follows. Let $V=Y \oplus Y^{*}$ be a symplectic vector space constructed from a vector space $Y$ of dimension $\geq 3$ over a field of two elements. Consider the group $G=V\rtimes \text{Sp}(V)$ where Sp$(V)$ is the group of symplectic linear transformations on $V$. $G$ is isocategorical but not isomorphic to a group $G_b$ which is an extension of Sp$(V)$ by $V$ using a cohomology class $b$. $V$ is the unique normal subgroup of $G$ and $G_b$ of order $|V|$ \cite{etingof2001isocategorical}. Therefore, the subgroup $V$ can be identified using the fusion of surface operators. Since $G$ is a semi-direct product while $G_b$ is a non-split group extension, using the results in section \ref{sec:groupext} it is clear that we can distinguish these two groups using surface operator fusion. 

\subsection{Groups with isomorphic surface operator fusion}
\label{sec:isosfusion}

In this section, we will derive necessary conditions for two groups to have isomorphic surface operator fusion and find a pair of such groups. 
Consider two groups $G_1$ and $G_2$ with the set of 2-representations 2Rep$(G_1)$ and 2Rep$(G_2)$. Consider the fusion of surface operators
\be
S_{(K_1,[c_1])} \times S_{(K_2,[c_2])}= \sum_{(K_3,[c_3])} N_{(K_1,[c_1]),(K_2,[c_2])}^{(K_3,[c_3])} S_{(K_3,[c_3])}~,
\ee
where $(K_i,[c_i])\in \text{2Rep}(G_1)$ and 
\be
S_{(H_1,[d_1])} \times S_{(H_2,[d_2])}=\sum_{(H_3,[d_3])} N_{(H_1,[d_1]),(H_2,[d_2])}^{(H_3,[d_3])} S_{(H_3,[d_3])}~,
\ee
where $(H_i,[d_i])\in \text{2Rep}(G_2)$. These fusions are said to be isomorphic if there is a bijective map 
\be
\Theta: \text{ 2Rep}(G_1)  \to \text{ 2Rep}(G_2)
\ee
such that 
\be
N_{\Theta((K_1,[c_1])),\Theta((K_2,[c_2]))}^{\Theta((K_3,[c_3]))}=N_{(K_1,[c_1]),(K_2,[c_2])}^{(K_3,[c_3])}~.
\ee
The existence of $\Theta$ is stricter than the isomorphism of the underlying twisted Burnside rings. Indeed, $\Theta$ shows that the twisted Burnside rings for the groups $G_1$ and $G_2$ are isomorphic and preserve the standard basis of simple 2-representations. 

Let us derive some properties of the map $\Theta$\footnote{For isomorphisms of fibered Burnside rings, similar results are proved in \cite{garcia2022species,boltje2022groups}}. Let $1_{G_i}$ be the trivial subgroup of $G_i$. Consider the fusion
\be 
S_{(1_{G_i},[1])} \times S_{(1_{G_i},[1])}= |G_i| ~ S_{(1_{G_i},[1])}~.
\ee
$S_{(1_{G_i},[1])}$ is the unique surface operator in 2Rep$(G_i)$ with this fusion rule. Since $\Theta$ preserves fusion rules, we have $\Theta((1_{G_1},[1]))=(1_{G_2},[1])$ and $|G_1|=|G_2|$. $\Theta$ should also map the trivial 2-representation $(G_1,[1])$ to the trivial 2-representation $(G_2,[1])$. 

Consider  $(K,[c])\in \text{2Rep}(G_1)$ and let $\Theta((K,[c]))=(H,[d])$ for some $(H,[d])\in \text{2Rep}(G_2)$. We will show that $|K|=|H|$. To prove this, consider the fusion
\be
S_{(K,[c])} \times S_{(1_{G_1},[1])}=  \frac{|G_1|}{|K|} S_{(1_{G_1},[1])}~.
\ee
Since $\Theta$ preserves fusion rules, the above fusion should be equal to
\be
S_{(H,[d])} \times S_{(1_{G_2},[1])}=  \frac{|G_2|}{|H|} S_{(1_{G_2},[1])}~.
\ee
Using $|G_1|=|G_2|$, we get $|K|=|H|$. Note that $G$ and $H$ need not be isomorphic as groups. 

Using this result we can show that if $\Theta((K,[c_1]))=(H_1,[d_1])$ and $\Theta((K,[c_2]))=(H_2,[c_2])$ then $H_1$ and $H_2$ are conjugate subgroups of $G_2$. To show this, consider the fusion rule
\be
S_{(K,[c_1])} \times S_{(K,[c_2])} = \sum_{g \in K\backslash G_1 /K} S_{K \cap \,\co{g}K,[c_1 \cdot \,\co{g}c_2]}~.
\ee
We have $N_{(K,[c_1]),(K,[c_2])}^{(K,[c_1 \cdot c_2])} \neq 0$. Since $\Theta$ preserves fusion rules, we have
\be
N_{(H_1,[d_1]),(H_2,[d_2])}^{(H_3,[d_3])} \neq 0~,
\ee
where $\Theta((K,[c_1 \cdot c_2]))=(H_3,d_3)$. This implies that there exists $g_1,g_2 \in G_2$ such that  $H_3=H_1 \cap \,\co{g}H_2$. Using $|K|=|H_1|=|H_2|=|H_3|$ we get $H_1=\co{g}H_2$. $\Theta$ therefore gives bijective maps between the sets $H^2(K,U(1))$ and $H^2(H,U(1))$. In general, this bijective map is not an isomorphism of cohomology groups. However, for subgroups with an additional property, this map can be shown to be an isomorphism. Suppose $K$ is a normal subgroup. We have the fusion rule
\be
S_{(K,[c_1])} \times S_{(K,[c_2])}= \sum_{g\in K\backslash G_1/K} S_{(K\cap \,\co{g}K,[c_1 \cdot \,\co{g}c_2])} = \sum_{g \in G_1/K} S_{(K,[c_1 \cdot \,\co{g}c_2])} ~.
\ee
The quotient group $G_1/K$ acts on $K$ by conjugation. This induces an action of $G_1/K$ on the cohomology group $H^2(K,U(1))$. If this action is trivial, then the fusion rule simplifies to
\be
\label{eq:H2iso1}
S_{(K,[c_1])} \times S_{(K,[c_2])}= \frac{|G_1|}{|K|} ~ S_{(K,[c_1 \cdot c_2])} ~.
\ee
Let $\Theta(K,[c_1])=(H,[d_1])$ and $\Theta(K,[c_2])=(H,[d_2])$. Note that $K$ being normal implies that $H$ is normal (since normality is determined by fusion rules). For the 2-representations $(H,[d_1])$ and $(H,d_2)$ we have the fusion
\be
\label{eq:H2iso2}
S_{(H,[d_1])} \times S_{(H,[d_2])}= \sum_{g\in H\backslash G_2/H} S_{(H\cap \,\co{g}H,[d_1 \cdot \,\co{g}d_2])} =  \sum_{g \in G_2/H} S_{(H,[d_1 \cdot \,\co{g}d_2])} ~.
\ee
Since $\Theta$ preserves fusion rules, comparing \eqref{eq:H2iso1} and \eqref{eq:H2iso2} we find that the R.H.S of \eqref{eq:H2iso2} should be of the form
\be
\frac{|G|}{|H|} ~ S_{(H,d)}
\ee
for some $d\in H^2(H,U(1))$. This is true if and only if $G_2/H$ acts trivially on the 2-cocycle $d_2$ and $d=d_1 \cdot d_2$. The fusion rule for $(H,[d_1])$ and $(H,d_2)$ simplifies to
\be
S_{(H,[d_1])} \times S_{(H,[d_2])}=\frac{|G_2|}{|H|} ~ S_{(H,[d_1 \cdot d_2])}~.
\ee
Since the cocycles $c_1$ and $c_2$ that we started with are arbitrary, we find that under the map $\Theta$ the cocycle $c_1 \cdot c_2$ gets mapped to $d_1 \cdot d_2$ for all $c_1,c_2\in H^2(K,U(1))$. Therefore, in this case $\Theta$ induces an isomorphism of groups $H^2(K,U(1))$
and $H^2(H,U(1))$. 

Consider  $(K,[c])\in \text{2Rep}(G_1)$, and let $\Theta((K,[c]))=(H,[d])$ for some $(H,[d])\in \text{2Rep}(G_2)$. We will show that if $[c]=[1]$, then $[d]=[1]$. To prove this, let $\Theta((K,[c])) =(H,[1])$ and $\Theta((K,[1]))=(H,[d])$. Consider the fusion
\be
S_{(K,[1])} \times S_{(K,[c])}= \sum_{g\in K\backslash G_1/K} S_{(K\cap \,\co{g}K,[\co{g}c])} ~.
\ee
We have $N_{(K,[1]),(K,[c])}^{(K,[c])} \neq 0$. Since $\Theta$ preserves fusion and using $\Theta((K,[1]))=(H,[d])$ we have 
\be
N_{(H,[d]),(H,[1])}^{(H,[1])} \neq 0~.
\ee
This shows that there exists $g \in G_2$ such that $[\co{g}d]=1$ which implies $[d]=[1]$. Therefore, the isomorphism $\Theta$ restricts to an isomorphism
\be
\Theta_B: B(G_1) \to B(G_2)
\ee
of Burnside rings of $G_1$ and $G_2$. Since this isomorphism preserves the standard basis of the Burnside rings, we get the following result:
\vspace{0.2cm}

{\it If the condensation surface defects for two groups $G_1$ and $G_2$ have isomorphic fusion rules, then the groups $G_1$ and $G_2$ have isomorphic table of marks.}
\vspace{0.2cm}

It is known that all groups of order $< 96$ have distinct table of marks \cite{kimmerle2005non}. Therefore, fusion of surface operators labelled by 2-representations uniquely determines any gauge group $G$ with $|G|<96$. Note that groups with isomorphic table of marks share many properties \cite{huerta2009some}.

\subsubsection{Examples}

In this section, we will construct two groups with isomorphic surface operator fusion. From the results in the previous section, we know that a necessary condition for two groups to have isomorphic surface operator fusion is that they should have isomorphic Burnside rings. In \cite{theevenaz1988isomorphic}, the author constructs an infinite family of groups which have isomorphic Burnside rings. Let us review this construction. 

Consider two prime numbers $p,q$ such that $q|(p-1)$ and $q\geq 3$. Let $a,b\in \DZ_{p}^{\times}$ of order $q$. Consider the cyclic groups
\be
Q= \DZ_q = \langle z \rangle, ~P_a= \DZ_p = \langle x \rangle, ~P_b= \DZ_p = \langle y~. \rangle, ~
\ee
Consider the action of $Q$ on $P_a$ and $P_b$ given by 
\be
z \cdot x= a x, ~ z \cdot y= b y~.
\ee
Let $G(a,b)$ be the semi-direct product $(P_a \times P_b) \rtimes Q$ constructed using this action. It is shown in \cite{theevenaz1988isomorphic} that the Burnside ring of $G(a,b)$ is independent of $a,b$. Also, it is shown that $G(a,b)\simeq G(c,d)$ if and only if $c=a^n,d=b^n$ for some $1\leq n < q$, and for a given $p,q$ there are $\frac{q-1}{2}$ isomorphism classes of groups with the same Burnside ring. The simplest example occurs for $p=1,q=5$ and groups $G(3,4)$ and $G(3,5)$. In the GAP Small Groups Library \cite{GAP4}, these groups can be accessed through the command SmallGroup(605,5) and SmallGroup(605,6). Though these two groups have the same Burnside ring, they can be distinguished using surface operator fusion. We know that the group of invertible surfaces is given by $H^2(G,U(1))$. Using GAP, it is easy to check that
\be
H^2(\text{SmallGroup}(605,5),U(1))=\DZ_1 ~, H^2(\text{SmallGroup}(605,5),U(1))=\DZ_{11}~.
\ee

For $p=29,q=7$ we get three non-isomorphic groups with isomorphic Burnside rings. These are groups SmallGroup(5887,5), SmallGroup(5887,6) and SmallGroup(5887,7) in GAP. We can compute the $2^{\text{nd}}$ cohomology groups for these three groups to find
\be
H^2(\text{SmallGroup}(5887,5),U(1))= \DZ_1 ~, H^2(\text{SmallGroup}(5887,6),U(1))= \DZ_{1}~,
\ee
\be
H^2(\text{SmallGroup}(5887,7),U(1))= \DZ_{29} ~.
\ee
Therefore, the fusion of invertible surface operators distinguishes the third group from the first two. What about the groups $G_1$:=SmallGroup(5887,5), $G_2$:=SmallGroup(5887,6)? These two groups have isomorphic surface operator fusion. To show this, let us look at the conjugacy classes of subgroups of these groups up to isomorphism. For  $G_1$, we get
\be
\label{eq:5Kexp1}
[ 1, 1 ], [ 7, 1 ], [ 29, 1 ], [ 29, 1 ], [ 29, 1 ], [ 29, 1 ], [ 29, 1 ], 
  [ 29, 1 ], [ 203, 1 ], [ 203, 1 ], [ 841, 2 ], [ 5887, 5 ] ~.
\ee
where the list contains the label of the groups in the Small Groups Library.  For $G_2$, we get
\be
\label{eq:5Kexp2}
 [ 1, 1 ], [ 7, 1 ], [ 29, 1 ], [ 29, 1 ], [ 29, 1 ], [ 29, 1 ], [ 29, 1 ], 
  [ 29, 1 ], [ 203, 1 ], [ 203, 1 ], [ 841, 2 ], [ 5887, 6 ] ~.
\ee
Therefore, the isomorphism class of conjugacy classes of subgroups of both $G_1$ and $G_2$ are the same except for the full group itself. We can compute the $2^{\text{nd}}$ cohomology group for these groups to find that $H^2(\text{SmallGroup(841,2)},U(1))= \DZ_{29}$ and the cohomology groups for all other subgroups are trivial. Therefore, the 2-representations of $G_1$ are labelled by the tuples 
\bea
\label{eq:G12reps}
&& (K_{1},[1]),(K_{7},[1]),(K^{(1)}_{29},[1]),(K^{(2)}_{29},[1]),(K^{(3)}_{29},[1]),(K^{(4)}_{29},[1]),(K^{(5)}_{29},[1]),(K^{(6)}_{29},[1]), \nonumber \\
&& (K^{(1)}_{203},[1]),(K^{(2)}_{203},[1]),(K_{841},[c]),(K_{5887},[5])~.
\eea
where the groups correspond to those in the list \eqref{eq:5Kexp1} and $[c] \in H^2(\text{SmallGroup(841,2)},U(1)) =\DZ_{29}$. Similarly, the 2-representations of $G_2$ are labelled by the tuples 
\bea
&& (H_{1},[1]),(H_{7},[1]),(H^{(1)}_{29},[1]),(H^{(2)}_{29},[1]),(H^{(3)}_{29},[1]),(H^{(4)}_{29},[1]),(H^{(5)}_{29},[1]),(H^{(6)}_{29},[1]), \nonumber \\
&& (H^{(1)}_{203},[1]),(H^{(2)}_{203},[1]),(H_{841},[d]),(H_{5887},[5])~.
\eea
where the groups correspond to those in the list \eqref{eq:5Kexp2} and $[d] \in H^2(\text{SmallGroup(841,2)},U(1)) =\DZ_{29}$.

We will show that the map 
\be
\Theta: \text{2Rep}(G_1) \to \text{2Rep}(G_2)
\ee
defined by 
\be
(K_i^{(n)},[l]) \to (H_i^{(n)},[l])
\ee
defines an isomorphism of surface operators preserving fusion rules. Since the groups $G_1$ and $G_2$ have isomorphic table of marks, the fusion coefficients satisfy
\be
N_{\trep_1, \trep_2}^{\trep_3}= N_{\Theta(\trep_1), \Theta(\trep_2)}^{\Theta(\trep_3)}~,
\ee
where $\trep_1,\trep_2,\trep_3$ are any 2-representations of $G_1$ in \eqref{eq:G12reps} with trivial $[c_{\trep_i}]$. In order to show that $\Theta$ is an isomorphism preserving fusion rules, we have to show that
\be
N_{(K_{841},[c]), \trep_2}^{\trep_3}= N_{\Theta((K_{841,[c]})), \Theta(\trep_2)}^{\Theta(\trep_3)}= N_{(H_{841,[d]}), \Theta(\trep_2)}^{\Theta(\trep_3)}
\ee
for any 2-representations $\trep_2,\trep_3$ of $G_1$. First, suppose $[c_{\trep_2}]=[1]$, then using \eqref{eq:fusioncoef} we have
\be
N_{(K_{841},[c]), \trep_2}^{\trep_3}= |\{ g\in K_{841}\backslash G_1/K_{\trep_2} ~|~ (K_{\trep_3},[c_{\trep_3}])=(K_{841} \cap \,\co{g}K_{\trep_2},[c])\}|
\ee
and 
\be
N_{(H_{841},[d]), \Theta(\trep_2)}^{\Theta(\trep_3)}= |\{ g\in H_{841}\backslash G_2/H_{\Theta(\trep_2)} ~|~ (H_{\Theta(\trep_3)},[d_{\Theta(\trep_3)}])=(K_{841} \cap \,\co{g}H_{\Theta(\trep_2)},[d])\}|~.
\ee
Since the groups $G_1$ and $G_2$ have the same table of marks
\be
|\{ g\in K_{841}\backslash G_1/K_{\trep_2} ~|~ K_{\trep_3}=K_{841} \cap \,\co{g}K_{\trep_2}\}|=|\{ g\in H_{841}\backslash G_2/H_{\Theta(\trep_2)} ~|~ H_{\Theta(\trep_3)}=K_{841} \cap \,\co{g}H_{\Theta(\trep_2)}\}|~.
\ee
Therefore, 
\be
N_{(K_{841},[c]), \trep_2}^{\trep_3}=N_{(H_{841},[d]), \Theta(\trep_2)}^{\Theta(\trep_3)}~,
\ee
where $[c_{\trep_2}]=[1]$. The only remaining case is to show the equality of the fusions 
\be
N_{(K_{841},[c_1]),(K_{841},[c_2])}^{(K_{841},[c_3])}= N_{(H_{841},[d_1]),(H_{841},[d_2])}^{(H_{841},[d_3])}~.
\ee
where $c_i \in H^2(K_{841},U(1))=\DZ_{29}$ and $d_i \in H^2(H_{841},U(1))=\DZ_{29}$. 
We have,
\bea
N_{(K_{841},[c_1]),(K_{841},[c_2])}^{(K_{841},[c_3])}&=& \{ g\in K_{841}\backslash G_1/K_{841} ~|~(K_{841},[c_3])=(K_{841}\cap \,\co{g}K_{841},[c_1 \cdot \,\co{g}c_2])\}| \nonumber \\
&=& |\{ g\in G_1/K_{841} ~|~  [c_3]= [c_1 \cdot \,\co{g}c_2])\}|~,
\eea
where we have used the fact that the groups $K_{841},H_{841}$ are normal subgroups of $G_1,G_2$, respectively. We have to understand the action of the group $G_1/K_{841}$ on the cohomology group $H^2(K_{841},U(1))=\DZ_{29}$. In the notation introduced at the beginning of this section, the group $K_{841}$ and $G_1/K_{841}$ are 
\be
K_{841}=\langle x,y\rangle \cong \DZ_{29} \times \DZ_{29} \text{ and } G_1/K_{841}= \langle z \rangle \cong \DZ_{29}~.
\ee 
The action of $z\in G_1/K_{841}$ on $K_{841}$ by conjugation is given by
\be
x\to x^{20}; ~ y \to  x^{23}~.
\ee
The 2-cocycle $c_k \in Z^2(K_{841},U(1))$ can be explicitly written as
\be
c_k((x^a,y^b),(x^c,y^d))=e^{\frac{2 \pi i k}{29} ad }~,
\ee
where $k \in \{0,...,28\}$. The action of $z\in G_1/K_{841}$ on $K_{841}$ by conjugation induces an action on the 2-cocycle given by
\be
c_k \to c_k^{25}~.
\ee
The orbit under the action of the full group $G_1/K_{841}$ on a 2-cocycle is 
\be
\label{eq:orb1}
c_k \to c_k^{25} \to c_k^{16} \to c_k^{23} \to c_k^{24} \to c_k^{20} \to c_k^{7} ~.
\ee

Similarly, we have to understand the action of the group $G_2/H_{841}$ on the cohomology group $H^2(H_{841},U(1))=\DZ_{29}$. In the notation introduced at the beginning of this section, the group $H_{841}$ and $G_2/H_{841}$ are 
\be
H_{841}=\langle x,y\rangle \cong \DZ_{29} \times \DZ_{29} \text{ and } G_2/H_{841}= \langle z \rangle \cong \DZ_{29}~.
\ee 
The action of $z\in G_2/H_{841}$ on $H_{841}$ by conjugation is given by
\be
x\to x^{20}; ~ y \to  x^{25}~.
\ee
The 2-cocycle $d_k \in Z^2(H_{841},U(1))$ can be explicitly written as
\be
d_k((x^a,y^b),(x^c,y^d))=e^{\frac{2 \pi i k}{29} ad }~.
\ee
where $k \in \{0,...,28\}$. The action of $z \in G_2/H_{841}$ on $H_{841}$ by conjugation induces an action on the 2-cocycle given by
\be
d_k \to d_k^{7}~.
\ee
The orbit under the action of the full group $G_2/H_{841}$ on a 2-cocycle is 
\be
\label{eq:orb2}
d_k \to d_k^{7} \to d_k^{20} \to d_k^{24} \to d_k^{23} \to d_k^{16} \to d_k^{25} ~.
\ee
Under the map $c_k \to d_k$ the set of 2-cocycles in the orbits \eqref{eq:orb1} and \eqref{eq:orb2} are the same. Therefore, under the map $\Theta$
\be
N_{(K_{841},[c_1]),(K_{841},[c_2])}^{(K_{841},[c_3])}= N_{(H_{841},[d_1]),(H_{841},[d_2])}^{(H_{841},[d_3])}
\ee
for all $c_i \in H^2(K_{841},U(1))=\DZ_{29}$ and $d_i \in H^2(H_{841},U(1))=\DZ_{29}$, as required.

This shows that surface operator fusion cannot determine a group up to isomorphism. It can be verified using GAP that the groups $G_1$:=SmallGroup(5887,5) and $G_2$:=SmallGroup(5887,6) have distinct character tables. Therefore, the fusion of Wilson lines \textit{can} distinguish these two groups. This shows that the character table cannot be constructed from the fusion rules of surface operators. Like the modular lattice of normal subgroups, many properties of the gauge group can be determined from either the fusion of Wilson lines or the fusion of surface operators. However, in general, the fusion of Wilson lines contains information about the gauge group that is not accessible from the fusion of surface operators and vice-versa. 

\section{$G$ from fusion of lines on surfaces}

\label{sec:GfromLonS}

So far we looked at the fusion of line and surface operators. More precisely, we looked at {\it genuine} line and surface operators whose description do not depend on higher dimensional manifolds. In this section, we will study the fusion of non-genuine line operators. In particular, we will study the fusion of line operators which ``live" on surface operators $S_{\rho}$. To set the stage for this discussion, let us start by studying junctions of Wilson lines. The set of local operators at a junction of Wilson lines $W_{R_1}$ and $W_{R_2}$ is given by the space of intertwiners (see for example \cite{bhardwaj2018finite})
\be
\text{Hom}(R_1,R_2)~.
\ee
From Schur's lemma, we know that for irreducible representations $R_1,R_2$ $\text{Hom}(R_1,R_2)=\delta_{R_1,R_2}\mathds{C}$. Therefore, two Wilson lines labelled by distinct irreducible representations cannot form a junction. 

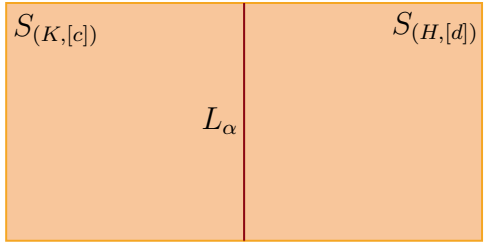
\begin{figure}[h!]
\centering

\tikzset{every picture/.style={line width=0.75pt}} 

\begin{tikzpicture}[x=0.75pt,y=0.75pt,yscale=-1,xscale=1]

\draw  [color={rgb, 255:red, 245; green, 166; blue, 35 }  ,draw opacity=1 ][fill={rgb, 255:red, 248; green, 198; blue, 155 }  ,fill opacity=1 ] (208,97) -- (328,97) -- (328,217) -- (208,217) -- cycle ;
\draw  [color={rgb, 255:red, 245; green, 166; blue, 35 }  ,draw opacity=1 ][fill={rgb, 255:red, 248; green, 198; blue, 155 }  ,fill opacity=1 ] (328,97) -- (448,97) -- (448,217) -- (328,217) -- cycle ;
\draw [color={rgb, 255:red, 139; green, 6; blue, 24 }  ,draw opacity=1 ]   (328,97) -- (328,217) ;

\draw (210,101.4) node [anchor=north west][inner sep=0.75pt]    {$S_{( K,[ c])}$};
\draw (401,100.4) node [anchor=north west][inner sep=0.75pt]    {$S_{( H,[ d])}$};
\draw (305,148.4) node [anchor=north west][inner sep=0.75pt]    {$L_{\alpha}$};

\end{tikzpicture}
\caption{Two surface operators $S_{(K,[c])}$ and $S_{(H,[d])}$ can form a junction and the line operators $L_{\alpha}$ at the junction are labelled by $\alpha \in \bigoplus_{g\in K\backslash G/H}\text{Rep}^{c \cdot \co{g}d}(K \cap \co{g}H)$.}
\label{fig:Latjunction}
\end{figure}

Two surface operators $S_{(K,[c])}$ and $S_{(H,[d])}$ can form a junction and the set of line operators at the junction is the set of morphisms \cite{bartsch2022non}
\be
\text{Hom}((K,[c]),(H,[d]))~.
\ee
between 2-representations $(K,[c])$ and $(K,[c])$ (see fig. \ref{fig:Latjunction}). This was computed in \cite{elgueta2004representation} to be
\be
\text{Hom}((K,[c]),(H,[d]))= \bigoplus_{g\in K\backslash G/H}\text{Rep}^{c \cdot \co{g}d}(K \cap \co{g}H)~,
\ee
where $\text{Rep}^{c \cdot \co{g}d}(K \cap \co{g}H)$ is the set of projective representations of $K \cap \co{g}H$ with 2-cocycle $c \cdot \co{g}d$. 

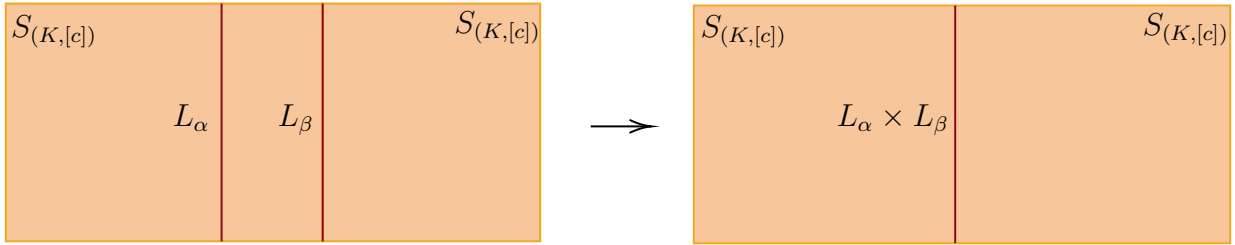
\begin{figure}[h!]
\centering

\tikzset{every picture/.style={line width=0.75pt}} 

\begin{tikzpicture}[x=0.75pt,y=0.75pt,yscale=-1,xscale=1]

\draw  [color={rgb, 255:red, 245; green, 166; blue, 35 }  ,draw opacity=1 ][fill={rgb, 255:red, 248; green, 198; blue, 155 }  ,fill opacity=1 ] (24,89) -- (183.83,89) -- (183.83,209) -- (24,209) -- cycle ;
\draw  [color={rgb, 255:red, 245; green, 166; blue, 35 }  ,draw opacity=1 ][fill={rgb, 255:red, 248; green, 198; blue, 155 }  ,fill opacity=1 ] (183.83,89) -- (293.5,89) -- (293.5,209) -- (183.83,209) -- cycle ;
\draw [color={rgb, 255:red, 139; green, 6; blue, 24 }  ,draw opacity=1 ]   (132.83,89) -- (132.83,209) ;
\draw [color={rgb, 255:red, 139; green, 6; blue, 24 }  ,draw opacity=1 ]   (183.83,89) -- (183.83,209) ;
\draw  [color={rgb, 255:red, 245; green, 166; blue, 35 }  ,draw opacity=1 ][fill={rgb, 255:red, 248; green, 198; blue, 155 }  ,fill opacity=1 ] (371,90) -- (641.5,90) -- (641.5,210) -- (371,210) -- cycle ;
\draw [color={rgb, 255:red, 139; green, 6; blue, 24 }  ,draw opacity=1 ]   (502.83,90) -- (502.83,210) ;
\draw    (319,151) -- (347.5,151) ;
\draw [shift={(349.5,151)}, rotate = 180] [color={rgb, 255:red, 0; green, 0; blue, 0 }  ][line width=0.75]    (10.93,-3.29) .. controls (6.95,-1.4) and (3.31,-0.3) .. (0,0) .. controls (3.31,0.3) and (6.95,1.4) .. (10.93,3.29)   ;

\draw (25.13,93.4) node [anchor=north west][inner sep=0.75pt]    {$S_{( K,[ c])}$};
\draw (248.87,91.4) node [anchor=north west][inner sep=0.75pt]    {$S_{( K,[ c])}$};
\draw (106.68,138.4) node [anchor=north west][inner sep=0.75pt]    {$L_{\alpha }{}$};
\draw (159.68,138.4) node [anchor=north west][inner sep=0.75pt]    {$L_{\beta }{}$};
\draw (441.68,138.4) node [anchor=north west][inner sep=0.75pt]    {$L_{\alpha } \times L_{\beta }{}$};
\draw (373,93.4) node [anchor=north west][inner sep=0.75pt]    {$S_{( K,[ c])}$};
\draw (596,92.4) node [anchor=north west][inner sep=0.75pt]    {$S_{( K,[ c])}$};

\end{tikzpicture}
\caption{Fusion of line operators on a surface is given by the composition of endomorphisms.}
\label{fig:compofL}
\end{figure}

The endomorphisms 
\be
\text{End}((K,[c]))=\text{Hom}((K,[c]),(K,[c]))= \bigoplus_{g\in K\backslash G/K}\text{Rep}^{d \cdot \co{g}d}(K \cap \co{g}K)
\ee
is the set of lines that ``live" on the surface $S_{(K,[c])}$. These line operators obey fusion rules given by the composition of endomorphisms (see fig. \ref{fig:compofL}). 
We are particularly interested in the case $K=\DZ_1$. Then we get
\be
\text{End}((\DZ_1,[1]))= \bigoplus_{g\in G}\text{Rep}(\DZ_1)~.
\ee
The line operators on the surface $S_{(\DZ_1,[1])}$ can be labelled as $L_g$ where $g\in G$ and their fusion is given by
\be
L_{g} \times L_h = L_{g \cdot h}~.
\ee
Therefore, it is clear that if we know the fusion of line operators on the surface $L_{(\DZ_1,[1])}$ we can reconstruct the group $G$ up to isomorphism. This is consistent with the expectation that the surface operators obtained from higher gauging a symmetry should contain line operators implementing the dual symmetry. In particular, the surface operator $S_{(\DZ_1,[1])}$ arises from higher gauging the full Rep$(G)$ symmetry. $S_{(\DZ_1,[1])}$ contains line operators implementing the dual symmetry of Rep$(G)$, which is precisely the group $G$. 

Note that even though the fusion of line operators on surface operators determines the gauge group, it is not a satisfactory answer to the reconstruction problem. This is because, thinking of $S_{(\DZ_1,[1])}$ as a trivial 2-dimensional QFT with $G$ 0-form symmetry, the results in this section can be paraphrased as the following statement: the gauge group of $\CT/G$ can be reconstructed from the fusion rules of line operators implementing the $G$-symmetry of $\CT$, which is obviously true, but a tautology. 

\section{Conclusion}

In this paper, we explored the fusion rules of Wilson lines and condensation surface operators obtained from higher-gauging Wilson lines in G-gauge theory, and the properties of the gauge group that can be derived from these fusion rules. We saw that the fusion of Wilson lines do not determine the group $G$ up to isomorphism. However, we can determine several properties of $G$ from them including the lattice of normal subgroups of $G$. We then looked at the fusion of surface operators obtained from condensing the Wilson lines on surfaces. We showed that the fusion of surface operators also determines the lattice of normal subgroups. Moreover, we can determine the order of the intersection of normal subgroups with all other subgroups. This allowed us to determine whether $G$ admits a presentation as a direct product, semi-direct product or a non-split group extension using the fusion rules. We  used this fact to show that infinite families of pairs of groups which cannot be distinguished using the fusion of Wilson loops can be distinguished using the fusion of surface operators. 

We defined isomorphism of fusion rules of surface operators for two groups. We derived several necessary conditions for such an isomorphism to exist. These necessary conditions pointed us to the pair of groups SmallGroup(5887,6) and SmallGroup(5887,7)  which have isomorphic fusion rules for surface operators. This pair of groups can be distinguished using fusion of Wilson lines which shows that the character table of a group cannot be determined from the fusion of surface operators. We learned that, in general, the fusion of Wilson lines contain information about the gauge group that is not accessible from the fusion of surface operators and vice-versa. Finally, we looked at the fusion of non-genuine line operators which live on surface operators and showed that their fusion rules determine the gauge group up to isomorphism. 

The discussion in this paper points to some natural questions as future directions:

\begin{itemize}
\item It will be interesting to find a pair of groups which have isomorphic fusion rules for Wilson lines and surface operators. Such a pair of groups will necessarily  have isomorphic character tables and table of marks. Groups satisfying these necessary conditions are known (see \cite{kimmerle2005non} and page 385 in \cite{lux1999computational}), and it would be interesting to determine whether these groups indeed have isomorphic Wilson line and surface operator fusion. 

\item An optimistic goal is to show that the group $G$ can be recovered up to isomorphism from the fusion rules of genuine operators. One may wonder why such a reconstruction should be possible. After all, the symmetric tensor category Rep$(G)$ which can be used to reconstruct $G$ seems to contain a lot of data. However, it is known that a group can be determined up to isomorphism using 1,2,3-characters of a group \cite{hoehnke19921}. These are generalizations of characters of a group defined by Frobenius \cite{frobenius1903primfactoren}. The higher representations of a group, whose tensor product structure govern the fusion rules of extended operators in a G-gauge theory, can also be used to define certain higher characters (as studied in Appendix \ref{Ax:2char}). Determining the relation between higher characters defined by Frobenius and higher characters arising from higher representations of a group is a potential way to determine whether the gauge group can be reconstructed from fusion rules of genuine operators.

\item Similar to the construction of 2-representations of $G$, we can construct $3$-representations by studying maps from the group to 3-vector spaces \cite{wang20153}. $3$-representations label a subset of membrane operators supported over $3$-dimensional submanifolds in a 4-dimensional $G$-gauge theory \cite{Bartsch:2022ytj}. $3$-representations are classified by the data $(K,[c])$ where $K$ is a subgroup of $G$ and $[c] \in H^3(K,U(1))$. This should be contrasted with 2-representations which are labelled by subgroups and their $2^{\text{nd}}$ cohomology group elements. The fusion rules for 3-representations (at least naively) contain additional data about the group not contained in the 2-representations. For example, the cohomology group $H^3(G,U(1))$ can be deduced from the tensor product of 3-representations of $G$. 

\item It will be interesting to determine a sufficient condition for two groups to have isomorphic surface operator fusion\footnote{See \cite{davydov2000finite} for a sufficient and necessary condition for two groups to have isomorphic character tables. In the context of fibred Burnside rings, a sufficient condition is proven in \cite{boltje2022groups}.}.

\item Even though the $S$ and $T$ matrices for Rep$(G)$ are trivial, it would be interesting to find and physically interpret invariants which capture the basis-independent data in the $F,R$ matrices\footnote{We thank Matthew Buican for suggesting this problem.}.

\item  It will be interesting to explore the properties of the gauge group contained in other line and surface operators apart from those considered in this paper. For example, the decomposition of a product of two conjugacy classes of a group into conjugacy classes can be determined from the fusion of flux lines in an untwisted discrete gauge theory. Also, to fully identify the gauging procedure, we need to also reconstruct the discrete torsion. This necessarily requires looking at non-Wilson lines. 

\item In this paper we focused on finite groups and associated extended operators. It will be interesting to determine the properties of a Lie group captured by fusion rules of operators. In the case of Lie groups, the Wilson lines are not topological in general. However, there are topological operators which label ``dual" symmetries arising from gauging Lie groups \cite{cheng2022gaugingLie,cheng2022gauging}.  
\end{itemize}

\ack{The author thanks ICTP for support. We thank Matthew Buican and Adrian Padellaro for numerous discussions, collaborations on related projects and detailed comments on a draft of this article. We thank Sanjaye Ramgoolam for collaboration on a related project.}

\begin{appendices}

\section{$G$ from 2-characters}

\label{Ax:2char}

\subsection{2-characters}

To define a character for 2-representations, we consider
\be
\tilde \chi_{\rho}(g):= \text{Tr}(\rho(g)):= \bigoplus_i \rho(g)_{ii}~.
\ee
$\tilde \chi_{\rho}(g)$ is valued in vector spaces. We would like a function valued in the complex numbers, which is invariant under conjugation of the entries. To define such a function, note that the isomorphisms $\phi_{g,h}$ imply the existence of the isomorphisms \cite{ganter2008representation}  
\be
\psi_{\rho}(g,h):= \text{Tr}(\rho(g)) \rightarrow \text{Tr}(\rho(hgh^{-1}))~.
\ee
When $g$ and $h$ commute, this is an endomorphism of Tr$(\rho(g))$. Therefore, for any two commuting elements $g,h \in G$, we can define the 2-character of the 2-representation $\rho(g)$ as
\be
\chi_{\rho}(g,h):= \text{Tr}(\psi_{\rho}(g,h))~.
\ee
$\psi_{\rho}(g,h)$ is a matrix valued in complex numbers. Therefore, the trace above is also valued in the complex numbers.  For any 2-representation labelled by a permutation $\sigma_n$ and a 2-cocycle $\tilde c$, the explicit expression for the characters was computed in \cite{osorno2010explicit} and is given by 
\be
\chi_{\rho}(g,h)= \sum_{i=\sigma_g(i)=\sigma_h(i)}  \frac{\tilde c_i(h,g^{-1})\tilde c_i(g,hg^{-1}) }{\tilde c_i(g,g^{-1})\tilde c_i(1,1)}~.
\ee
The 2-characters satisfy
\be
\chi_{\rho_1 \times \rho_2}= \chi_{\rho_1} \chi_{\rho_2}, ~~
\chi_{\rho_1 + \rho_2}=\chi_{\rho_1} + \chi_{\rho_2}~.
\ee
Unlike characters of 1-representations of a group, distinct 2-representations can have the same 2-character \cite{osorno2010explicit}.  

Since we are interested in the irreducible 2-representations labelled by $(K,[c])$, it is convenient to write the expression for the character in terms of this data. This can be done using the expression for the induced character of a 2-representation in \cite{ganter2008representation} to get
\be
\label{eq:2char}
\chi_{\rho}(g,h)= \frac{1}{|K|} \sum_{s\in G, sgs^{-1}\in K, shs^{-1}\in K}  \frac{c(shs^{-1},sg^{-1}s^{-1}) c(sgs^{-1},shg^{-1}s^{-1}) }{\tilde c_i(sgs^{-1},sg^{-1}s^{-1})}~.
\ee
Since the 2-representation $\rho$ is determined by the label $(H,[c])$, we will denote this 2-character as $\chi_{(H,[c])}(g,h)$ in the following sections. 

\subsection{Normal Subgroups from 2-characters}

Suppose the group $G$ has a normal subgroup $N$, then the 2-character for the 2-representation $(N,[1])$, where $[1]$ is the trivial element of the cohomology group $H^2(N,U(1))$ is given by
\be
\chi_{(N,[1])}(g,h)= \frac{1}{|N|} \sum_{s\in G, sgs^{-1}\in N, shs^{-1}\in N}  1~.
\ee
Since $N$ is normal, $sgs^{-1},shs^{-1}\in N$ if and only if $g,h \in N$. Moreover, if $g,h \in N$, then $sgs^{-1},shs^{-1}\in N$ for all $s \in G$. Therefore, the 2-character is given by
\be
\chi_{(N,[1])}(g,h)= \begin{cases}
\frac{|G|}{|N|} & \text{ if } g,h \in N ~,\\
0 & \text{otherwise} ~.
\end{cases}
\ee
In particular for the normal subgroups $\DZ_1,G$ we get the 2-characters
\bea
\chi_{(\DZ_1,[1])}(g,h)&=& \begin{cases}
\frac{|G|}{|N|} & \text{ if } g=h=1 ~,\\
0 & \text{otherwise}~,
\end{cases} \\
\chi_{(G,[1])}(g,h)&=& 1 ~ \forall ~ g,h \in G~.
\eea

Conversely, if we have the 2-character $\chi_{(K,[c])}(g,h)$ of some 2-representation $(H,[c])$, can we determine whether $H$ is a normal subgroup? To answer this question, consider 
\bea
|\chi_{(K,[c])}(g,h)|&=& \bigg |\frac{1}{|K|} \sum_{s\in G, sgs^{-1}\in K, shs^{-1}\in K}  \frac{c(shs^{-1},sg^{-1}s^{-1}) c(sgs^{-1},shg^{-1}s^{-1}) }{\tilde c_i(sgs^{-1},sg^{-1}s^{-1})}\bigg | \cr
 &\leq&  \frac{1}{|K|} \sum_{s\in G, sgs^{-1}\in K, shs^{-1}\in K}  \bigg | \frac{c(shs^{-1},sg^{-1}s^{-1}) c(sgs^{-1},shg^{-1}s^{-1}) }{\tilde c_i(sgs^{-1},sg^{-1}s^{-1})}\bigg | \cr
 &=& \frac{1}{|K|} \sum_{s\in G, sgs^{-1}\in K, shs^{-1}\in K}  1 \leq \frac{|G|}{|K|}~.
\eea
In particular, if $|\chi_{(K,[c])}(g,h)|=\frac{|G|}{|K|}$, then
\be
sgs^{-1} \in K , shs^{-1} \in K ~ \forall ~ s \in G~.
\ee
If $K$ is not normal, then there exists some $g',h',s' \in G$ such that $s'g's'^{-1},s'h's'^{-1} \notin K$ and consequently $|\chi_{(K,[c])}(g',h')|<\frac{|G|}{|K|}$. 

\textit{In summary,
\be
\label{eq:normal}
|\chi_{(K,[1])}(g,h)|= \begin{cases}
\frac{|G|}{|K|} & \text{ if } g,h \in K ~,\\
0 & \text{otherwise}~,
\end{cases}
\ee
if and only if $K$ is a normal subgroup of $G$. Moreover, if $K$ is not normal, then $\chi_{(K,[1])}(g,0)\neq 0 <\frac{|G|}{|K|}$ for some $g \notin K$}.

Therefore, by inspecting the 2-characters we can determine whether the group $G$ has a normal subgroup, and if so, the order of the normal subgroup is given by $\frac{|G|}{|\chi_{(K,[c])}(1,1)|}$.

\subsection{2-character for $G/N$}

In section \ref{sec:quotientfus}, we studied 2-representations of $G/N$ and learned that they correspond to a subset of 2-representations of $G$  of the form $(K,[c])$ where $K$ is a subgroup of $G$ containing $N$ and $[c] \in H^2(K,U(1))$ is the inflation of some $[r] \in H^2(K/N,U(1))$. The character of the 2-representation $(K/N,[r])$ of $G/N$ is given by
\be
\chi_{(K/N,[r])}(k,l)= \chi_{(K,[c])}(g,h)~,
\ee
where $g$ and $h$ satisfies $\pi(g)=k, \pi(h)=l$. The choice of $g,h$ does not matter. Indeed, consider $n_1g$ and $n_2h$ where $n_1,n_2\in N$ and $g,h\in G$ such that $\pi(n_1g)=k$ and $\pi(n_2h)=l$. $k$ and $l$ commutes in $G/N$ iff $n_1g$ and $n_2h$ commutes in $G$ for all $n_1,n_2\in N$. Consider
\be
\chi_{(K,[c])}(n_1g,n_2h)= \frac{1}{|K|} \sum_{s\in G, sn_1gs^{-1}\in K, sn_2hs^{-1}\in K}  \frac{c(sn_2hs^{-1},s(n_1g)^{-1}s^{-1}) c(sn_1gs^{-1},shg^{-1}s^{-1}) }{\tilde c_i(sn_1gs^{-1},s(n_1g)^{-1}s^{-1})}~.
\ee
If $sn_1gs^{-1} \in K$ then $sn_1s^{-1}sgs^{-1}\in K$ which implies that $sgs^{-1}\in K$ (note that $n_1,n_2 \in N \subseteq K$). Also, since $c=r\circ \pi$, $c(sgs^{-1},shs^{-1})=c(sn_1gs^{-1},sn_2hs^{-1})$. Therefore, we have 
\be
\chi_{(K/N,[r])}(k,l)= \chi_{(K,[c])}(n_1g,n_2h)=\chi_{(K,[c])}(g,h)~, 
\ee
for all $n_1,n_2 \in N$. 

\subsection{Groups extensions and 2-characters}

In section \ref{sec:groupext}, we deduced several special fusion rules for surface operators which tell us about the group extension structure of the group $G$. In particular, we studied the fusion rule \eqref{eq:dpfusions} which implies that $G$ is a direct product. Using the properties of 2-characters for normal subgroups that we studied above, we can rephrase \eqref{eq:dpfusions} in terms of 2-characters. $G$ is a direct product of normal subgroups of $K$ and $N$ if and only if 
\be
\label{eq:dpchar}
|\chi_{(K,[1])}(g,h)|= \begin{cases}
\frac{|G|}{|K|} & \text{ if } g,h \in K ~,\\
0 & \text{otherwise}~,
\end{cases} \quad |\chi_{(N,[1])}(g,h)|= \begin{cases}
\frac{|G|}{|N|} & \text{ if } g,h \in N ~,\\
0 & \text{otherwise}~,
\end{cases}
\ee
and 
\be
\chi_{(K,[1])}(g,h) \cdot \chi_{(N,[1])}(g,h)= \chi_{(\DZ_1,[1])}(g,h)~.
\ee
The conditions for $G$ to be a semi-direct product or a non-split group extension (see \eqref{eq:sdpfusions}  and following discussion) can again be rephrased in terms of characters. $G$ is a semi-direct product of subgroups $K$ and $N \triangleleft G$ if and only if 
\be
|\chi_{(N,[1])}(g,h)|= \begin{cases}
\frac{|G|}{|N|} & \text{ if } g,h \in N ~,\\
0 & \text{otherwise}~,
\end{cases}
\ee
and 
\be
\chi_{(K,[1])}(g,h) \cdot \chi_{(N,[1])}(g,h)= \chi_{(\DZ_1,[1])}(g,h)~,
\ee
and $K$ doesn't satisfy \eqref{eq:dpchar}. Finally, if there are no 2-characters of $G$ satisfying the above conditions, $G$ can only be presented as a non-split group extension. 

\end{appendices}

\newpage

\bibliography{bibfile}

\end{document}